\documentclass[a4paper,11pt]{article}
\pdfoutput=1 

\usepackage{jheppub} 

\usepackage[T1]{fontenc} 
\usepackage{textcomp}
\usepackage{units}

\newcommand{\pt}{p_{\perp}}

\title{Medium response in JEWEL and its impact on jet shape observables in heavy ion collisions}

\author[a]{Raghav Kunnawalkam Elayavalli}
\author[b,c]{and Korinna Christine Zapp}

\affiliation[a]{Rutgers, The State University of New Jersey, Piscataway, New Jersey 08854, USA}
\affiliation[b]{Laborat\'{o}rio de Instrumenta\c{c}\~{a}o e F\'{\i}sica Experimental de Part\'{\i}culas (LIP), Av. Prof. Gama Pinto, \textnumero 2, 1649-003 Lisboa, Portugal}
\affiliation[c]{Theory Department, CERN, CH-1211 Gen\`eve 23, Switzerland}

\emailAdd{raghav.k.e@cern.ch}
\emailAdd{korinna.zapp@cern.ch}

\abstract{
	Realistic modeling of medium-jet interactions in heavy ion collisions is becoming increasingly important to successfully predict jet structure and shape observables. In \textsc{Jewel}, all partons belonging to the parton showers initiated by hard scattered partons undergo collisions with thermal partons from the medium, leading to both elastic and radiative energy loss. The recoiling medium partons carry away energy and momentum from the jet. Since the thermal component of these recoils' momenta is part of the soft background activity, comparison with data requires the implementation of a  subtraction procedure. We present two independent procedures through which background subtraction can be performed and discuss the impact of the medium recoil on jet shape observables. Keeping track of the medium response significantly improves the \textsc{Jewel} description of jet shape measurements.
}

\begin{document} 
\maketitle
\flushbottom

\section{Introduction}

Hard probes in heavy ion collisions that traverse the hot and dense medium, often referred to as the quark gluon plasma (QGP), undergo significant energy loss. One of the avenues to study this energy loss is by the use of reconstructed jets. Results from experiments at the LHC, have shown consistent findings regarding the degree to which the jet spectra is quenched~\cite{Aad:2014bxa,Adam:2015ewa,Khachatryan:2016jfl} and the recovery of the ``lost" energy at large angles from the jet~\cite{Khachatryan:2015lha}. Simultaneously, medium induced modification to the jet structure have also been observed in these experiments. These results show an increase in the multiplicity of low $\pt$ jet constituents as seen in the jet fragmentation function~\cite{Chatrchyan:2014ava,Aad:2014wha,ATLAS-CONF-2015-055}, and that these constituents are found in the periphery of the jet~\cite{Chatrchyan:2013kwa}. An increase of relatively soft particle multiplicity is also observed at large angles from the jet axis~\cite{Khachatryan:2016tfj}. At the same time the hard component of jets gets narrower, as observed in a slight reduction of the jet girth~\cite{Cunqueiro:2015dmx}. Recent results have also probed the so called groomed sub-jet shared momentum fraction~\cite{CMS:2016jys}, where Pb+Pb jets are split more asymmetrically than p+p jets in the lower $\pt$ bins, with the effect disappearing at higher $\pt$. The invariant charged jet mass~\cite{Acharya:2017goa} was also recently measured in different jet $\pt$ bins and found to be largely unmodified as compared to p+p. Together these results point to a picture of modifications to the jet structure in heavy ion collisions that involve a narrowing of the jet core along with a broadening of the jet at its periphery. 

\smallskip

It has been observed~\cite{Milhano:2015mng,Rajagopal:2016uip,Casalderrey-Solana:2016jvj} that the narrowing of the jet core, accompanied by a hardening of the fragmentation function at large $z$, is due to the fact that broader jets with a softer fragmentation pattern are more susceptible to energy loss and are thus more likely to fail analysis cuts and disappear from the sample. The increase of soft particle production at relatively large angles from the jet axis is commonly attributed to the medium's response to the energy and momentum deposited by the jet~\cite{Casalderrey-Solana:2016jvj,KunnawalkamElayavalli:2016dda,Tachibana:2017syd}, although other interpretations exist~\cite{Chien:2015hda}.

Since the latest high luminosity runs at the LHC, the focus in heavy ion jet studies has moved to detailed characterisation of jet shapes and intra-jet observables to highlight the modification to the internal structure of the jet. The impact that medium response has on such jet shape or jet sub-structure observables thus offers a possibility to observe the thermalisation of energy and momentum deposited by a jet. In this context, the advantage of jet observables is that the additional soft activity can be clearly identified inside a jet, while globally it is much more difficult to separate it from the much larger, soft, fluctuating background.

Theoretically, medium response is described using different frameworks. At strong coupling it has been known for some time that energy and momentum lost by a quark is transferred into hydrodynamic modes and leads to a Mach cone and wake~\cite{Gubser:2007ga,Chesler:2007sv} (for reviews see e.g.~\cite{Chesler:2015lsa,CasalderreySolana:2011us}). Medium response at weak coupling has also been studied in a hydrodynamic framework~\cite{Qin:2009uh,Neufeld:2009ep}, supported by the observation of fast thermalisation of soft fragments~\cite{Iancu:2015uja}. While these approaches rely on a clear separation between jet and medium degrees of freedom, transport codes treat all partons, whether they are hard or soft, on equal footing. In these calculations, soft partons that interacted with a jet, undergo further re-scattering and thereby thermalise~\cite{Bouras:2014rea,He:2015pra,Gao:2016ldo}. Recently, hybrid approaches have been developed, that describe the propagation of jets in transport theory, but treat the thermal medium and its response to energy and momentum depositions by jets in hydrodynamics~\cite{Casalderrey-Solana:2016jvj,Chen:2017zte,Tachibana:2017syd}. A different kind of hybrid approach is followed by~\cite{Casalderrey-Solana:2016jvj}, where the energy lost by a weak coupling jet is calculated at strong coupling and the medium response is again treated in hydrodynamics.

In the jet event generator \textsc{Jewel}~\cite{Zapp:2013vla} three options for dealing with medium response are available: (i) one can ignore it, (ii) extract a source term for a hydrodynamic treatment~\cite{Floerchinger:2014yqa}, or (iii) keep thermal partons recoiling against interactions with the jet in the event and let them hadronise together with the jet. In this paper, we explore the latter option. It is organised as follows: In section~\ref{sec::jewelrecoils} we discuss the handling of medium response in the \textsc{Jewel} generator, followed by a description of two subtraction techniques in section~\ref{sec::submethods}. After a brief specification of the used Monte Carlo samples in section~\ref{sec::samples} we then characterise the subtraction methods systematically in section~\ref{sec::systematics}. We then discuss the effects of medium response on traditional jet quenching observables (section~\ref{sec::tradobs}) and jet shape observables (section~\ref{sec::jetshapes}) and close with a discussion of our results in section~\ref{sec::conclusions}.

\section{Treatment of medium response in JEWEL}	
\label{sec::jewelrecoils}

In \textsc{Jewel} the background medium is assumed to consist of an ensemble of partons, the phase space distribution and flavour composition of which have to be provided by an external medium model. Partons belonging to a jet may interact with these background partons through $2\to 2$ scattering processes described by perturbative matrix elements, with associated gluon emission generated by the parton shower. Further details of the inner workings and Monte Carlo implementation of \textsc{Jewel} are available in~\cite{Zapp:2012ak,Zapp:2011ya}.

As mentioned in the introduction, there are two operational modes for event generation with \textsc{Jewel} concerning the treatment of background partons recoiling against a scattering with the jet (so called ``recoils'' or ``recoiling partons''). Events can be generated with or without storing the recoil information. When run without recoils, the recoiling partons do not show up in the event. In this case no medium response is considered and inclusive and inter-jet observables can be compared to (background subtracted) experimental data. So far this was the recommended mode for jet observables. 

However, jet structure observables are sensitive to medium response and hence it is desirable to include these effects in \textsc{Jewel} by keeping the recoiling partons in the event. After the scattering these recoiling partons do not interact further with the medium and free-stream towards hadronisation. This represents a limiting case for the recoil behavior, that can be regarded as being the limit opposite the assumption of immediate thermalisation of recoil energy and momentum made by hydrodynamic frameworks. The truth is expected to be between these two extreme cases, since one would expect that these partons interact further with the medium, but do not necessarily fully thermalise. 

So far the background partons could be either (anti-)quarks or gluons. For hadronisation, however, all recoiling partons are converted to gluons. It is assumed that the recoiling parton is a colour neighbour of the hard parton it interacted with. The recoil gluons are thus inserted in the strings connecting the partons forming the jet. Therefore, the hadronic final state including recoiling partons is not an incoherent superposition of jets and activity arising from recoils. At hadron level, it is impossible to assign a certain hadron to the jet or medium response.

The four-momentum of the recoiling partons has two components: the thermal momentum it had before the interaction with the jet, and the four-momentum transferred from the jet in the scattering process. Only the latter is interesting for investigating medium response, the former is part of the uncorrelated thermal background that is subtracted from the jet. As \textsc{Jewel} generates only the jets and not full heavy ions events, it is not possible to use the experimental background subtraction techniques for the Monte Carlo events. Instead, a dedicated procedure for removing the thermal four-momentum components from the jets when running with recoils has to be devised. Therefore, along with the recoiling partons, we are also storing the thermal four-momenta, which constitute our background\footnote{Technically, this is done by adding one line labeled as comment for each thermal momentum to be subtracted to the HepMC event record.}. These will be systematically removed from the jets during the analysis step, as detailed in the following section.

\section{Subtraction of the thermal component}
\label{sec::submethods}

As discussed in the previous section, in order to compare predictions for jet  observables with data, it is imperative to perform a background subtraction on \textsc{Jewel} events when running with the recoils. This is to avoid a mismatch between the prediction and data, since the jets in data have the fluctuating underlying event subtracted. In this section, we present two independent subtraction methods for \textsc{Jewel}, that can be employed at the analysis level\footnote{Example \textsc{Rivet}~\cite{Buckley:2010ar} analyses are available for download on the \textsc{Jewel} homepage \url{http://jewel.hepforge.org/}}. 

\subsection{4MomSub}

This method removes the thermal momenta exactly from the jet's four-momentum. In order to determine which thermal momenta should be subtracted, an additional set of neutral particles with very small energy and momenta and pointing in the direction of the thermal momenta are added to the final state particles list. These ``dummy" particles are effectively the same as ghosts that \textsc{FastJet}~\cite{Cacciari:2011ma} uses during its clustering to determine the jet area. They can get clustered into jets without affecting the jet's momentum or structure. Thermal momenta, that are matched to a dummy (in the azimuthal angle - pseudorapidity plane) inside a jet, are subtracted from the jet's momentum. The resulting four vector constitutes the subtracted jet momentum. An algorithmic implementation of the procedure is detailed below: 
\begin{enumerate}
	\item Cluster the initial jet collection from the final state particles (including dummies). 
	\item Compile a list of the thermal momenta (particles in the HepMC event record with status code 3).
	\item For each jet, get the list of thermal momenta that have $\Delta R < 1\cdot 10^{-5}$ with one of the jet constituents, i.e a dummy particle.
	\item Sum up the four-momenta of the matched thermal momenta. This constitutes the background.
	\item For each jet subtract the background four-momentum from the jet's four momentum, this provides the corrected jet collection.
	\item Calculate jet observables from corrected jet four-momenta. 
\end{enumerate}

This method is easily generalised to subtraction of sub-systems of jets, such as sub-jets or annuli used for the jet profile.

\subsection{GridSub}
\label{sec::gridsub}

This is a generic, observable independent subtraction method. A finite resolution grid (in the $\phi-\eta$ plane) is superimposed on the jet and its constituents. The four-momenta of particles in each cell in the grid are then vectorially summed and thermal momenta subtracted, yielding the cell four-momentum. Finally, we re-cluster the jet with the cell four-momenta as input to the jet clustering algorithm. This method does not require dummy particles. It is also possible to first discretise the entire event, subtract thermal four-momenta cell-by-cell, and then cluster jets. The algorithms for the two variants are given below.

\smallskip

\textbf{Jet clustering before discretisation (GridSub1):}
\begin{enumerate}
	\item Cluster the initial jet collection from the final state particles. 
	\item Compile a list of the thermal momenta (particles in the HepMC event record with status code 3).
	\item Define the grid resolution and place grid over jets.
	\item Inside each cell sum the jet constituents' four-momenta and subtract the thermal four-momenta that fall into the cell (note: no matching is required, thermal four-momenta with distance $\Delta R < R$ from the jet axis are considered\footnote{Alternatively, when dummy particles are written to the event record, one can also match thermal momenta and dummies to decide which momenta should be included.}), providing a single four momentum for each cell.
	\item In case a cell contains more thermal momentum than jet constituents, the cells is set to have zero four-momentum. This is deemed to be the case when the (scalar) $\pt$ of the thermal component is larger than the $\pt$ of the particle component.
	\item Re-cluster the jets with the cell four-momenta as input to get the final, subtracted jets.
	\item Calculate jet observables from re-clustered jets. 	
\end{enumerate}
This version is the default.

\smallskip

\textbf{Discretisation before jet clustering (GridSub2):}
\begin{enumerate}
	\item Compile a list of the thermal momenta (particles in the HepMC event record with status code 3).
	\item Define the grid resolution and place grid over the entire event.
	\item Inside each cell sum the final state particles' four-momenta and subtract the thermal four-momenta that fall into the cell (note: no matching is required), providing a single four momentum for each cell.
	\item In case a cell contains more thermal momentum than particle momentum, the cells is set to have zero four-momentum. This is deemed to be the case when the (scalar) $\pt$ of the thermal component is larger than the $\pt$ of the particle component.
	\item Cluster the jets with the cell four-momenta as input to get the final, subtracted jet.
	\item Calculate jet observables. 	
\end{enumerate}

Due to the finite size of the grid, it is possible to have certain cells with more thermal momentum than particle momentum, resulting in a total negative four-momentum, which in our case is set to zero before clustering. Thus, the GridSub method systematically removes less background from the jet than 4MomSub. The smearing introduced by the GridSub method will be quantified systematically in the following section.

The use of the 4MomSub method is recommended when possible, since it does not introduce finite-resolution effects and is consequently more accurate.

\subsection{Limitations of the subtraction and the issue of track jets}
\label{sec:limitations}

Since the subtraction techniques introduced above subtract the thermal momenta, which are at parton level, from the hadronic final state, they only yield meaningful results for observables that are insensitive to hadronisation effects. This is the case for most infra-red safe observables based on calorimetric jets. Examples for observables that do not fall into this category are fragmentation functions and charged jet observables. In general, all cuts on the final state particles, also $\pt$ cuts, are problematic.

A few of the recent experimental results involve the use of charged or track jets~\cite{Abelev:2013kqa,Cunqueiro:2015dmx,Acharya:2017goa}, i.e\ jets reconstructed using only tracks. When the subtraction is naively applied,  the techniques end up overestimating the contribution of the four-momenta to subtract. Thus, in order to compare with such experimental results, a heuristic procedure is applied. The observable of interest is calculated for full jets and re-scaled. The re-scaling between the full and the charged jet distribution is extracted from the corresponding \textsc{Jewel} simulation for p+p collisions.  If it is larger than the resolution of the observable, it is applied to the full jet subtracted distribution in Pb+Pb. In this way an estimate of the charged jet distribution is derived. For example, a naive way of estimating the charged jet four-momentum is by re-scaling the full jet quantity with the fraction of charged particles in the jet. The charged jet mass distribution discussed in section~\ref{sec::jetshapes} is estimated using this technique and compared with data. In other cases, for instance the jet radial moment girth (also shown in  section~\ref{sec::jetshapes}), the distributions for charged and full jets are the same in p+p collisions. In this case we compute the observable for full jets in Pb+Pb collisions as well and do not apply any re-scaling.

Obviously, this method comes with an additional uncontrolled systematic uncertainty, since it is not guaranteed that the same relation between full and charged jet distribution holds in Pb+Pb and p+p.

\section{The event sample}
\label{sec::samples}

We generate di-jet events in the standard setup~\cite{Zapp:2013vla} at $\sqrt{s_\text{NN}} = \unit[2.76]{TeV}$ and $\sqrt{s_\text{NN}} = \unit[5.02]{TeV}$ with the simple parametrization of the background discussed in detail in~\cite{Zapp:2013zya}. The values used for the formation time and initial temperature,  $\tau_\text{i}=\unit[0.6]{fm}$ and  $T_\text{i}=\unit[485]{MeV}$ for $\sqrt{s_\text{NN}} =  \unit[2.76]{TeV}$ and $\tau_\text{i}=\unit[0.4]{fm}$ and $T_\text{i}=\unit[590]{MeV}$ for $\sqrt{s_\text{NN}} = \unit[5.02]{TeV}$, are taken from a hydrodynamic calculation~\cite{Shen:2012vn,Shen:2014vra}. The proton PDF set is \textsc{Cteq6LL}~\cite{Pumplin:2002vw} and for the Pb+Pb sample the \textsc{Eps09}~\cite{Eskola:2009uj} nuclear PDF set is used in addition, both are provided by \textsc{Lhapdf}~\cite{Whalley:2005nh}. 

We use the \textsc{Rivet} analysis framework~\cite{Buckley:2010ar} for all our studies. Jets are reconstructed using the same jet algorithm as the experiments (anti-$k_\perp$~\cite{Cacciari:2008gp}) from the \textsc{FastJet} package~\cite{Cacciari:2011ma}.

For our studies we use Pb+Pb samples with and without medium response and a corresponding p+p sample. In addition, we also need samples at parton level for some of the systematic studies, as they allow for a parton-by-parton separation of jet and recoil contributions.

\section{Systematic studies}
\label{sec::systematics}

The background subtraction techniques introduced in the previous section and their effects on jets are studied henceforth in a systematic fashion. 

\subsection{Smearing due to finite resolution of the grid}

\begin{figure*}[h] 
	\centering
	\includegraphics[width=0.47\textwidth]{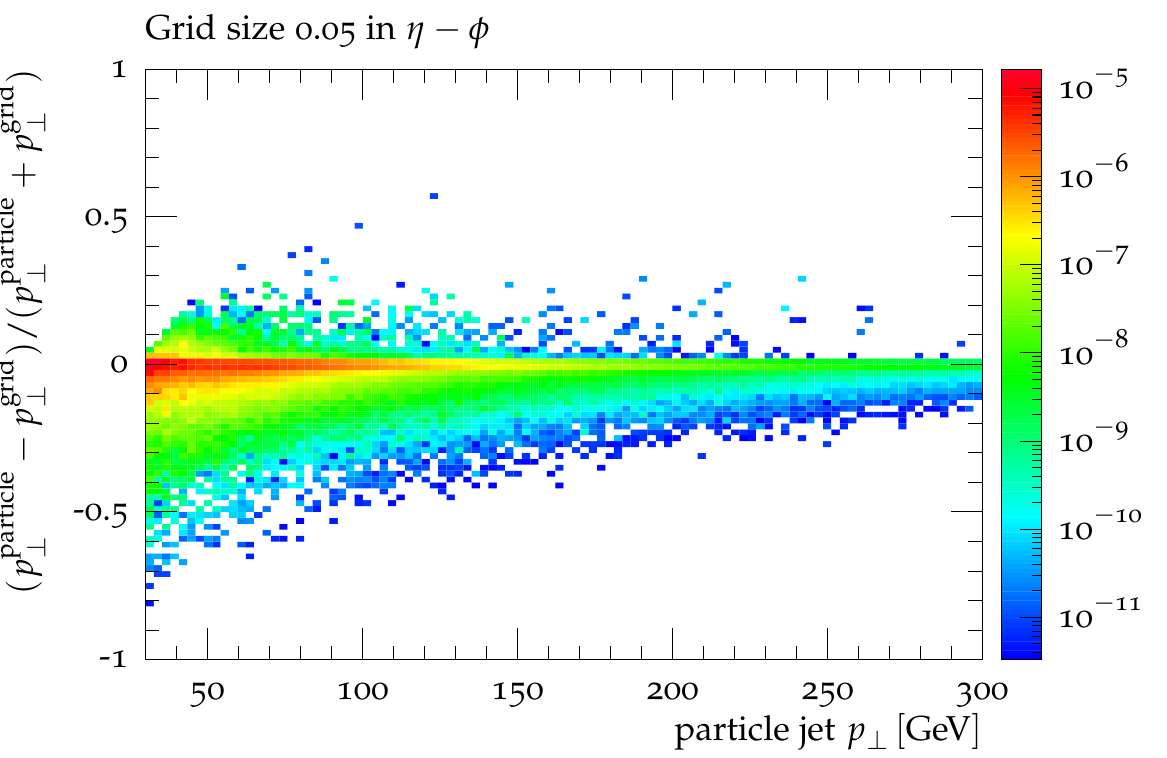} 
	\includegraphics[width=0.47\textwidth]{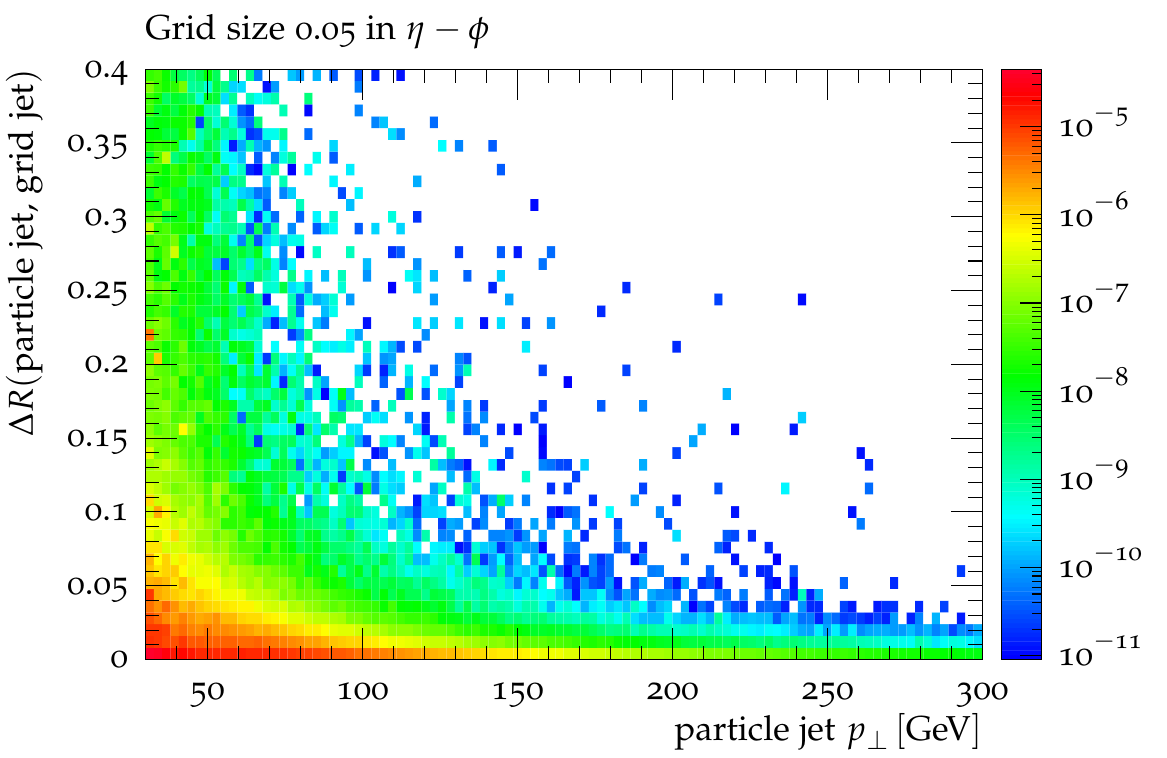} 
	\includegraphics[width=0.47\textwidth]{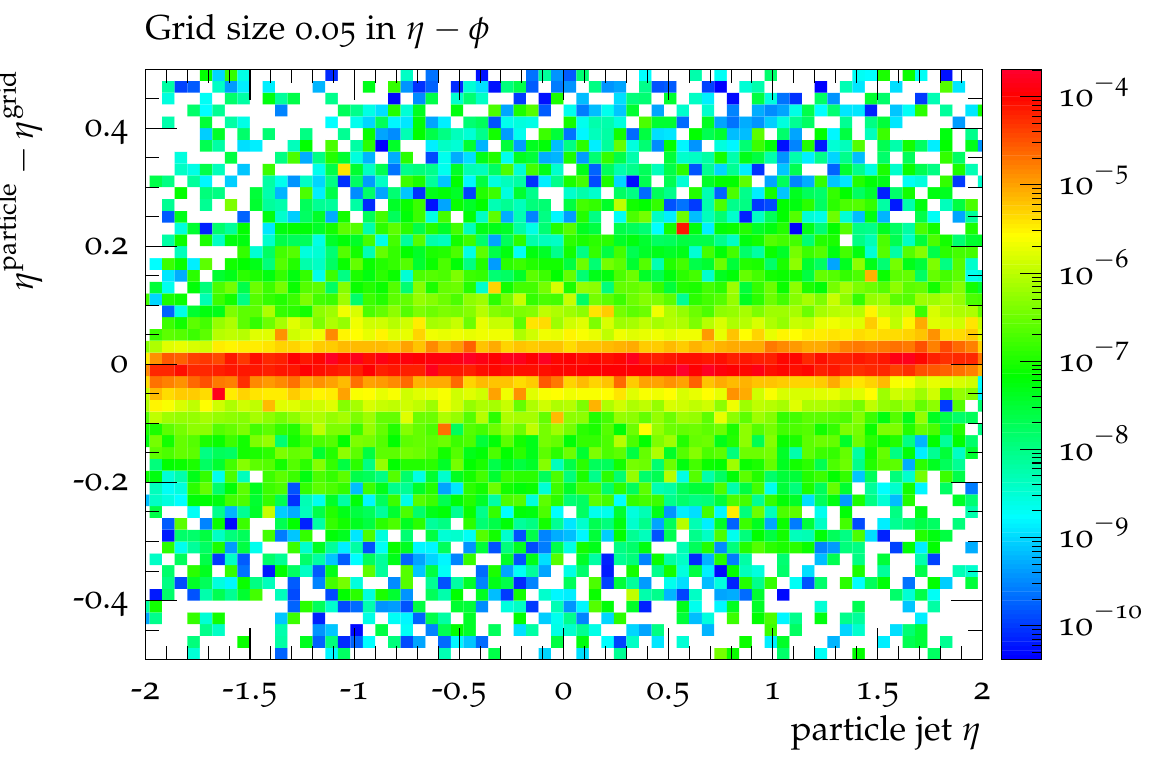} 
	\includegraphics[width=0.47\textwidth]{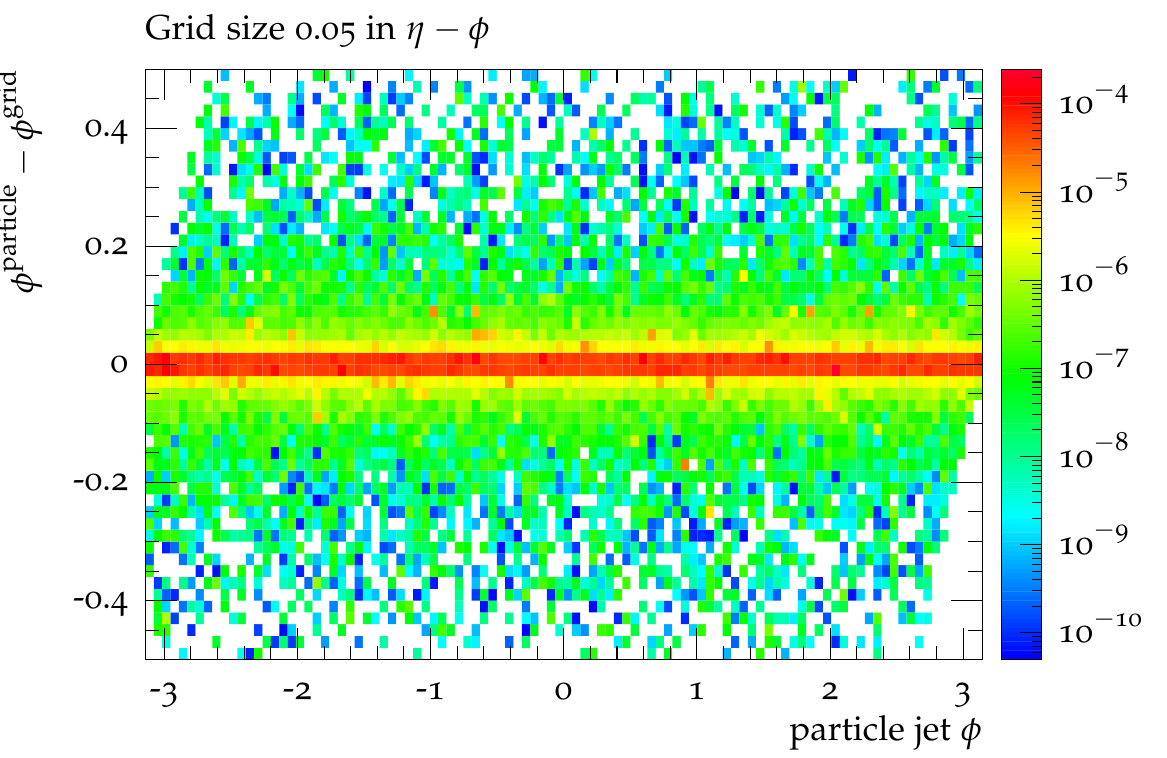} 
	\caption{Smearing introduced by the grid on the jets, quantized by the smearing in jet $\pt$ (top left) and the absolute shift of the jet axis in the $\eta-\phi$ plane (top right) shown as a function of the particle jet $\pt$. Bottom plots show the relative shifts in jet $\eta$ (left) and $\phi$ (right), shown as a function of the respective particle jet $\eta$ and $\phi$. (Note that the log scales on the z axis span six orders of magnitude.)}
	\label{fig:gridjetsmearing}
\end{figure*}

An immediate consequence of the grid, before any subtraction is introduced, is that both the jet $\pt$ and the position of the jet in the $\eta-\phi$ plane are smeared. This effect is studied in \textsc{Jewel} with p+p events generated at hadron level to highlight the inherent behavior. All our studies of the grid are shown for a nominal grid size of $0.05$ in $\eta-\phi$ plane, which we find to be a good compromise between resolution and under-subtraction (which is more severe for smaller cell sizes). The systematic uncertainties are estimated by varying the grid size by a factor of two and most final observables are shown to be quite insensitive to the grid size within these limits.

In each event, jets are first reconstructed from the final state hadrons. Then the event is discretised using a grid and jets are reconstructed based on the grid cells. Finally, each jet of the smaller of the two collections is matched to the one from the other set that is closest in $\Delta R$, with the constraint that $\Delta R$ is smaller than the reconstruction radius (this is the standard CMS procedure for comparing generator level jets to jets after detector simulation). The smearing is quantified in Fig.~\ref{fig:gridjetsmearing} with the top panels showing the smearing in jet $\pt$ (on the left) and jet axis (on the right) as a function of the particle jet $\pt$. The latter is broken down into the respective shifts in $\eta$ and $\phi$, which are shown in the bottom panels. The deviations are observed to be small in the $\pt$ range studied here. There is a clear trend for the grid jet $\pt$ to be larger than the corresponding particle jet $\pt$, which is due to the fact that the effective area of the grid jets can be larger due to the discretisation. As one would expect, increasing the jet $\pt$ reduces the smearing introduced by the grid. 

\begin{figure}[h] 
	\centering
	\includegraphics[width=0.47\textwidth]{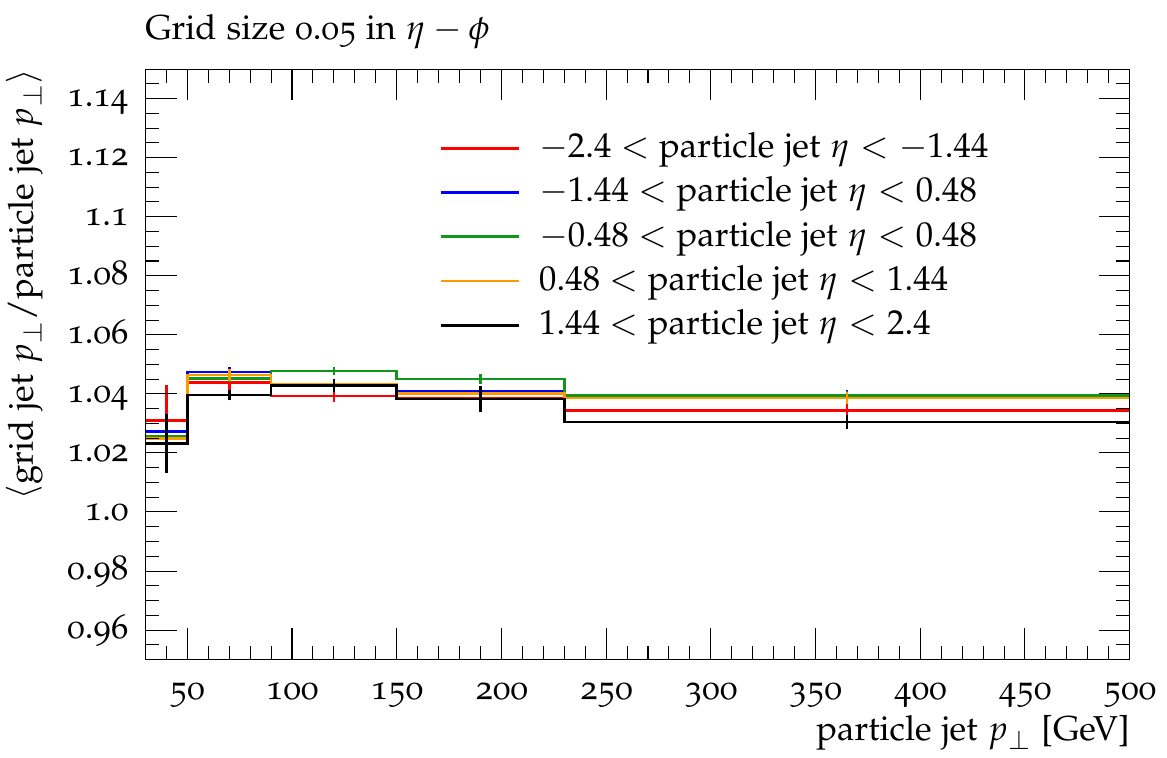} 
	\includegraphics[width=0.47\textwidth]{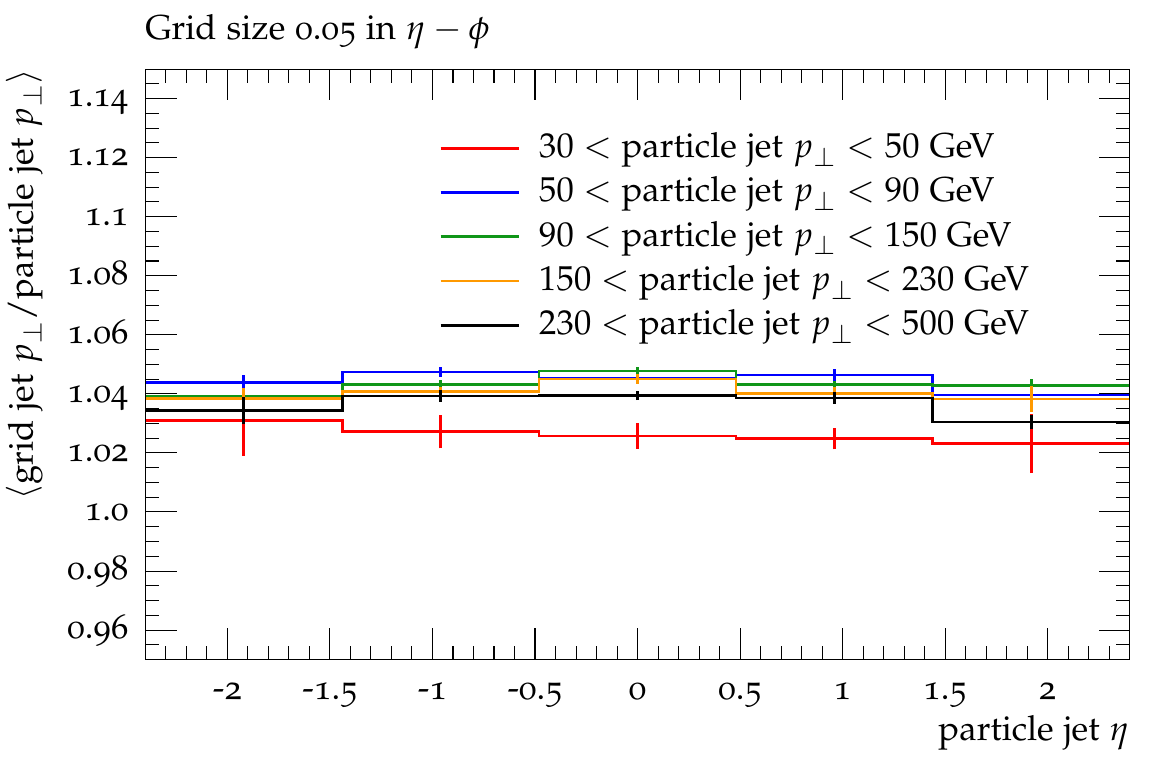} 
	\caption{Average shift in the jet $\pt$ introduced by the discretisation as a function of the particle jet $\pt$ and $\eta$, respectively.}
	\label{fig:jes}
\end{figure}		

In Fig.~\ref{fig:jes} the $\pt$ shift seen in Fig.~\ref{fig:gridjetsmearing} is quantified. The ratio between grid jet $\pt$ and the particle jet $\pt$ is seen to be around 1.04 and thus reasonably close to unity, and largely independent of jet $\pt$ and $\eta$ for $\pt^\text{jet} > \unit[50]{GeV}$. Such shifts usually are corrected in experiments~\cite{Chatrchyan:2011ds,Berta:2016ukt} by introducing detector level correction factors as a function of the jet $\pt$ and $\eta$. In this paper, GridSub jets are not corrected for this shift in their $\pt$, since it is reasonably small. Also, it partially cancels when looking at ratios of Pb+Pb with p+p due to its independence on jet kinematics. Furthermore, since the mismatch is related to nearby jets, increasing the jet $\pt$ cut leads to a reduction of the effect.

\subsection{Under-subtraction due to cells with negative energy}

\begin{figure}[h] 
	\centering
	\includegraphics[width=0.47\textwidth]{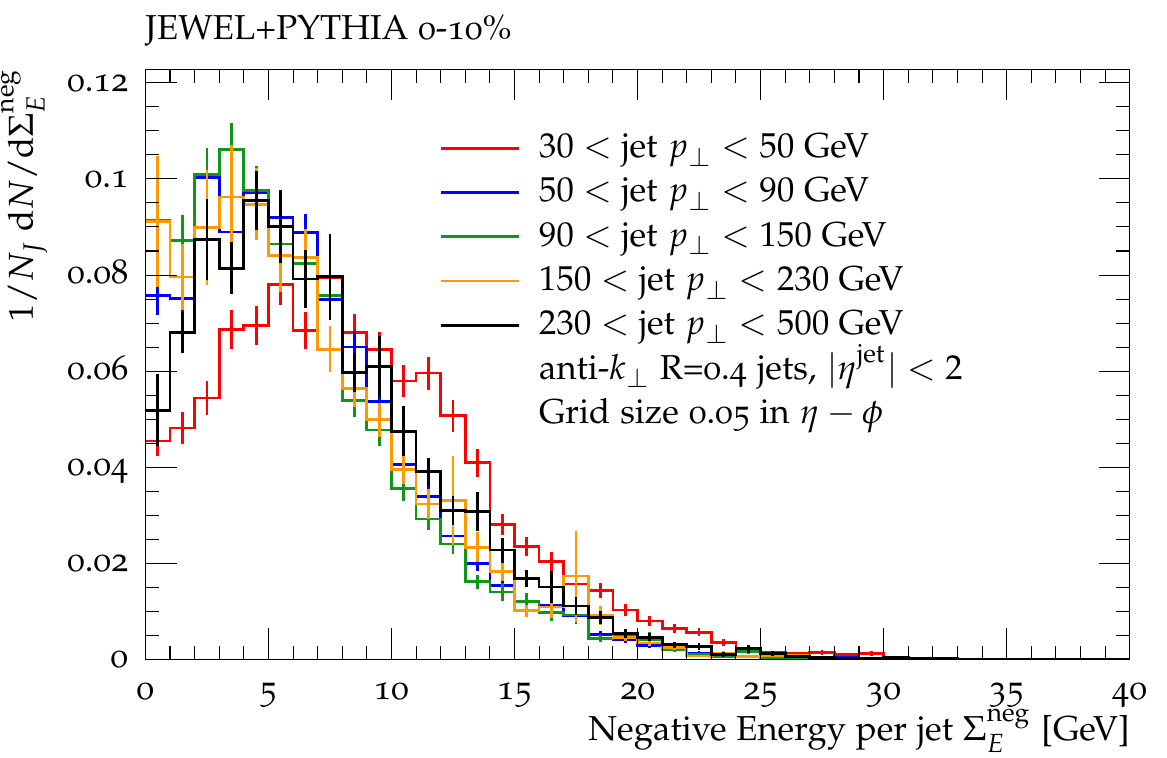} 
	\caption{Total negative energy per jet introduced by the GridSub technique shown for different jet $\pt$ bins  for central Pb+Pb ($0-10\%$) \textsc{Jewel+Pythia} events generated with recoils.}
	\label{fig:gridnegativeenergy}
\end{figure}		

As previously mentioned, the GridSub technique sets the cell's four momentum to zero if it contains more thermal than particle momentum. This leads to a systematic under-subtraction, that increases with decreasing cell size. We quantify this effect using the event sample with medium response included. Jets are reconstructed and subtracted using the default grid subtraction, but here we keep track of the energy of cells whose four-momentum is set to zero. For each jet we then check if it contains such cells and sum the (negative) energy that these cells originally had. The sum of the negative energy per grid jet is shown in Fig.~\ref{fig:gridnegativeenergy} for different jet $\pt$ ranges. The contribution of negative energy, i.e the amount of thermal energy that remains un-subtracted from the jet, is largely independent of the jet $\pt$ (except for the lowest $\pt$ bin) and small compared to the jet $\pt$ over most of the covered $\pt$ range. 

\subsection{Comparison of two GridSub versions}

\begin{figure*}[h] 
	\centering
	\includegraphics[width=0.47\textwidth]{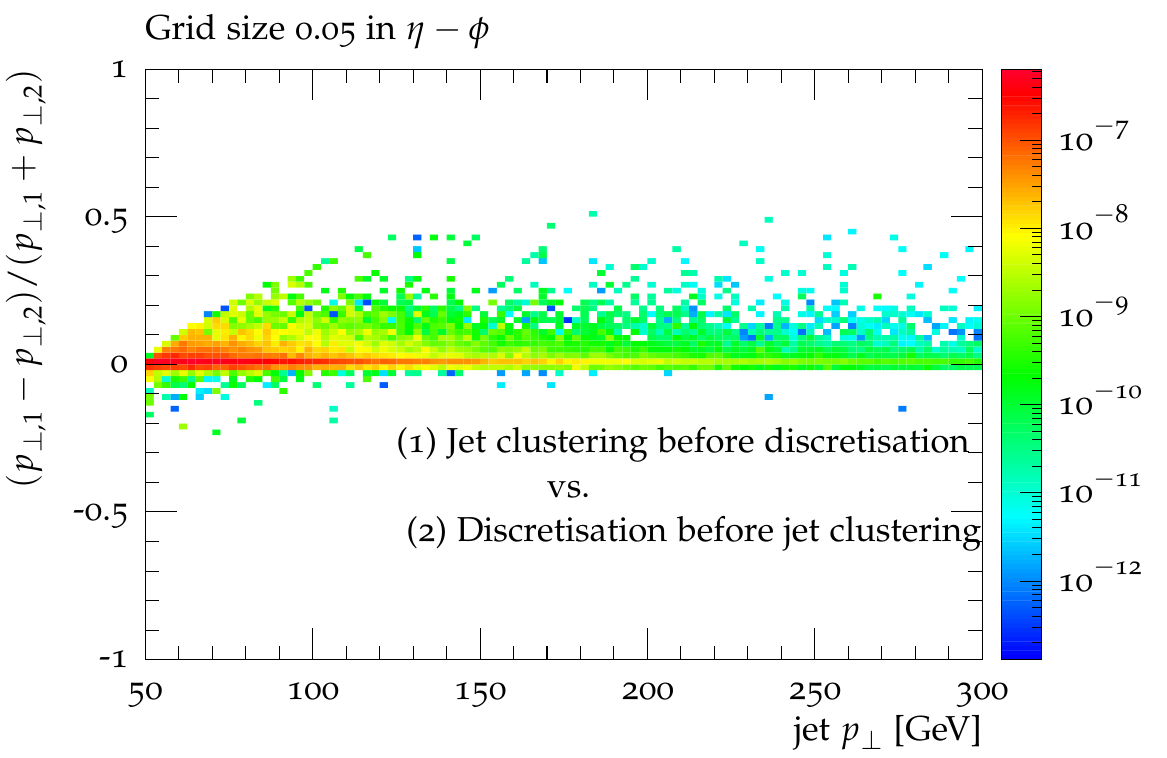} 
	\includegraphics[width=0.47\textwidth]{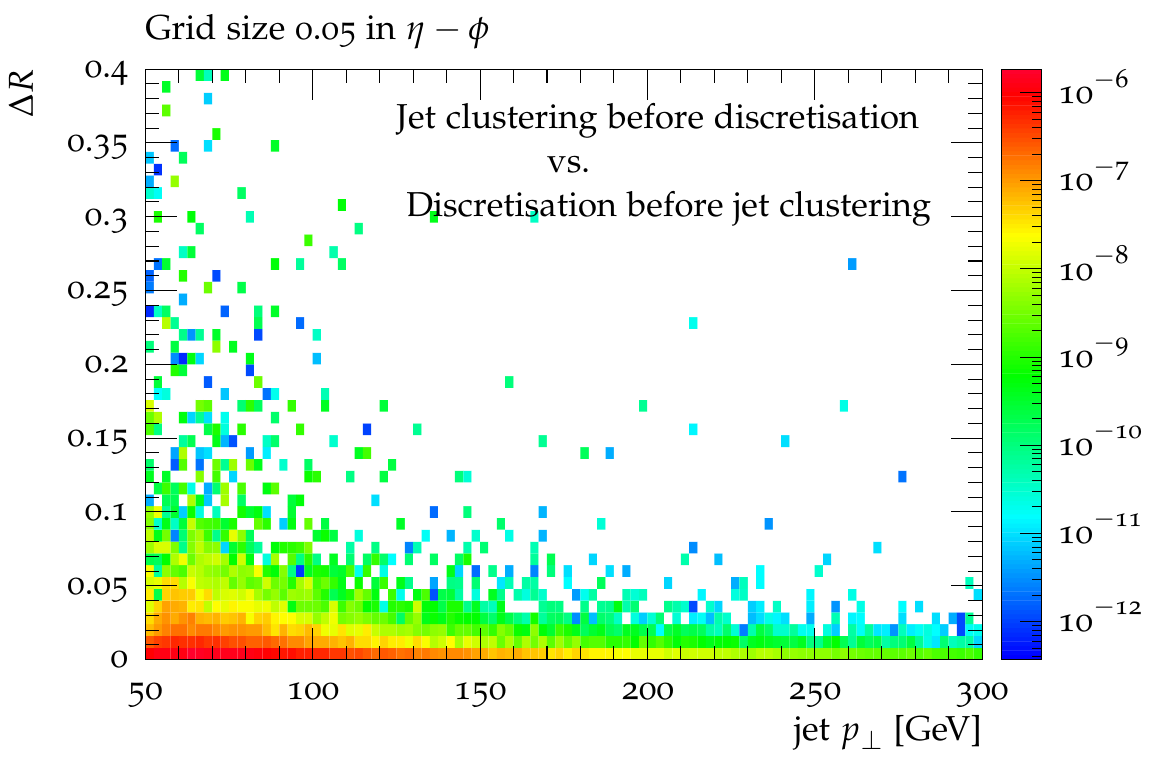} 
	\caption{Relative $\pt$ difference (left) and shift of jet axis (right) between the two GridSub versions ((1) jet clustering before discretisation and (2) discretisation before jet clustering).}
	\label{fig:asybkgsubjetfinding}
\end{figure*}

As discussed in section~\ref{sec::gridsub} we have implemented two versions of the grid based subtraction, that differ in the order of jet clustering and discretisation. It is to be expected that the two versions yield different results, as there is no reason why the two operations should commute. Using again the hadron level event sample with medium response included we quantify the differences between the versions. To this end, we find and subtract jets with both versions and event-by-event match the jets following the procedure detailed above. The relative difference in jet $\pt$ and shift of the axis due to the different ordering of operations is shown in Fig.~\ref{fig:asybkgsubjetfinding}. Both these effects are determined to be quite small, but the jet $\pt$ is consistently larger, when the initial jet clustering is performed before the discretisation of the event.

\subsection{Effects on jet $\pt$ with 4MomSub and GridSub subtraction}

\begin{figure}[h] 
	\centering
	\includegraphics[width=0.47\textwidth]{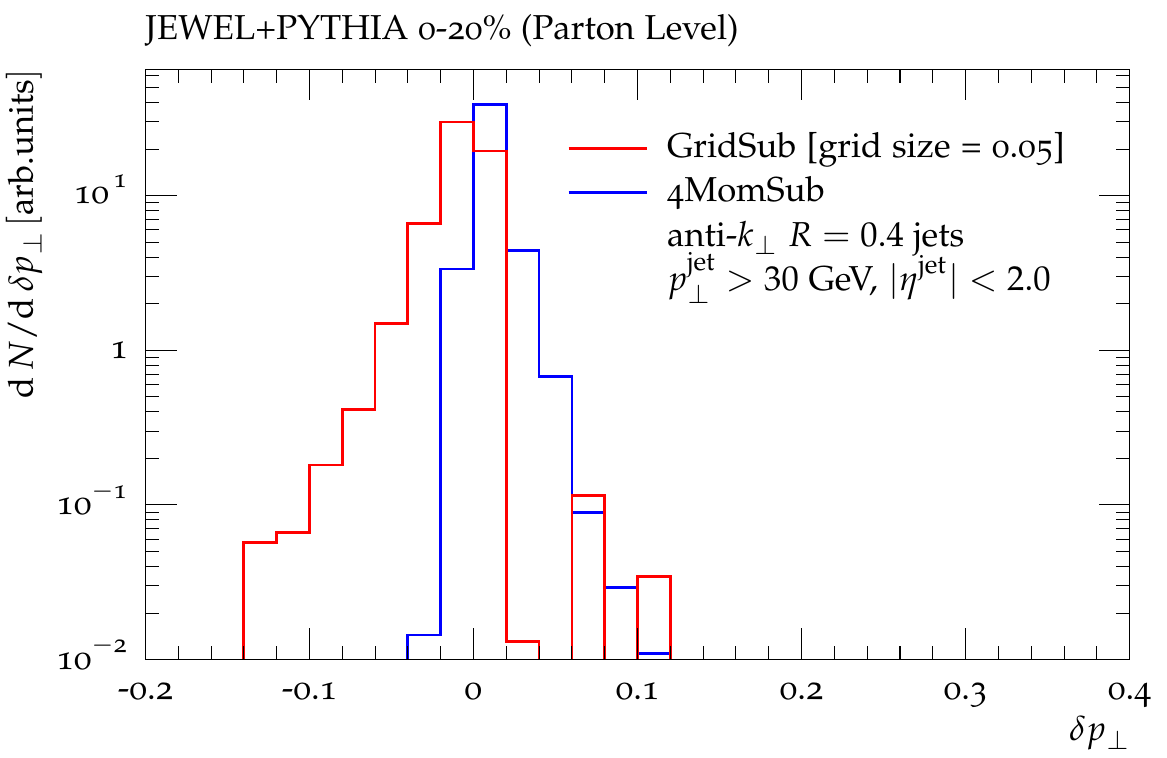} 
	\caption{Relative $\pt$ difference $\delta \pt = (\pt^\text{w/o\ rec} - \pt^\text{w\ rec})/(\pt^\text{w/o\ rec} + \pt^\text{w\ rec})$ between parton level jets reconstructed with and without recoiling partons for the two subtraction methods.}
	\label{fig:subAsym_partonLevel}
\end{figure}

A final check of the two subtraction methods (4MomSub and GridSub1) is done at parton level, where the same jets can be reconstructed with and without recoiling partons. The subtraction is performed with either of the two methods and the matching procedure is again the same as before. 

Fig.~\ref{fig:subAsym_partonLevel} shows the relative $\pt$ difference between jets reconstructed with and without recoiling partons. As expected, the 4MomSub distribution is narrower compared to GridSub1, due to additional jet smearing introduced by the discretization of the event into cells of a finite size. Additionally, the 4MomSub distribution has a tail on the positive side. This is a momentum conservation effect: the thermal distribution is isotropic (except for the longitudinal boost), while the recoiling partons have a net momentum in direction of the jet due to momentum conservation. Therefore, when including medium response more momentum is added to the jet than is subtracted. This is a physical effect that is independent of the subtraction method, but for the GridSub method the shoulder is towards the negative side. This is due to the aforementioned nature of the GridSub to under-subtract the jets, which overcompensates the momentum conservation effect.  

\section{Application to traditional jet quenching observables}	
\label{sec::tradobs}	

\begin{figure}[h] 
	\centering
	\includegraphics[width=0.47\textwidth]{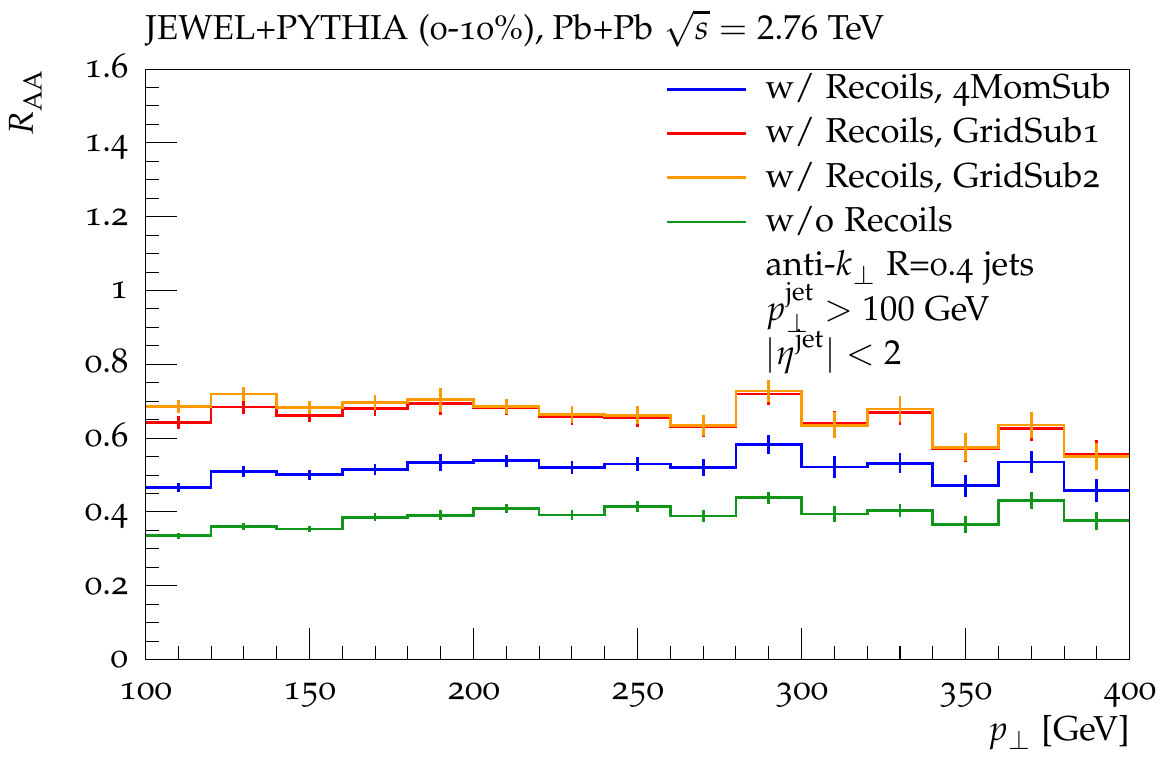} 
	\caption{Inclusive jet nuclear modification factor $R_\text{AA}$ for Pb+Pb central events in \textsc{Jewel+Pythia}. The green line represents \textsc{Jewel+Pythia} without medium response while the blue, red and orange lines show the result including medium response with 4MomSub, GridSub1 and GridSub2 respectively.}
	\label{fig:effect_bkgsub}
\end{figure}		

Observables built from the jet $\pt$ and axis, such as jet $R_{\rm{AA}}$ or the di-jet asymmetry $A_J$, for smaller radii jets typically show a rather mild sensitivity to medium response. The jet axis is dominated by the hard jet components and for the jet $\pt$ the only effect of medium response is a partial recovery of lost energy. For small reconstruction radii, this is at best a moderate effect, while for very large radii, such as $R \approx 1.0$, the effect becomes sizable. For such large radii also the systematic uncertainties related to the subtraction become large. Experimentally, the study of such large jets in a heavy ion environment constitutes an almost impossible task of discriminating between underlying event and the jets. For small radii jets at small momenta the same problem persists, which is why different experiments utilize different procedures to remove the effect of the underlying event in the jet collection of interest~\cite{Adam:2015ewa,Aaboud:2017bzv,Kodolova:2007hd}.  

As our primary validation, Fig.~\ref{fig:effect_bkgsub} shows the nuclear modification factor $R_{\rm{AA}}$ of jets, i.e the ratio of jet yield in Pb+Pb over binary collisions scaled p+p, for a moderate radius of $R=0.4$. As expected, including medium response leads to a small increase of $R_{\rm{AA}}$ over the entire jet $\pt$ range. The grid based subtraction leads to a significantly larger increase. This reflects the under-subtraction of the GridSub method discussed in section~\ref{sec::systematics}. Increasing the cell size leads to a reduction of $R_{\rm{AA}}$. There is good agreement between the two versions of the grid subtraction.

\begin{figure}[t]
	\centering
	\includegraphics[width=0.47\textwidth]{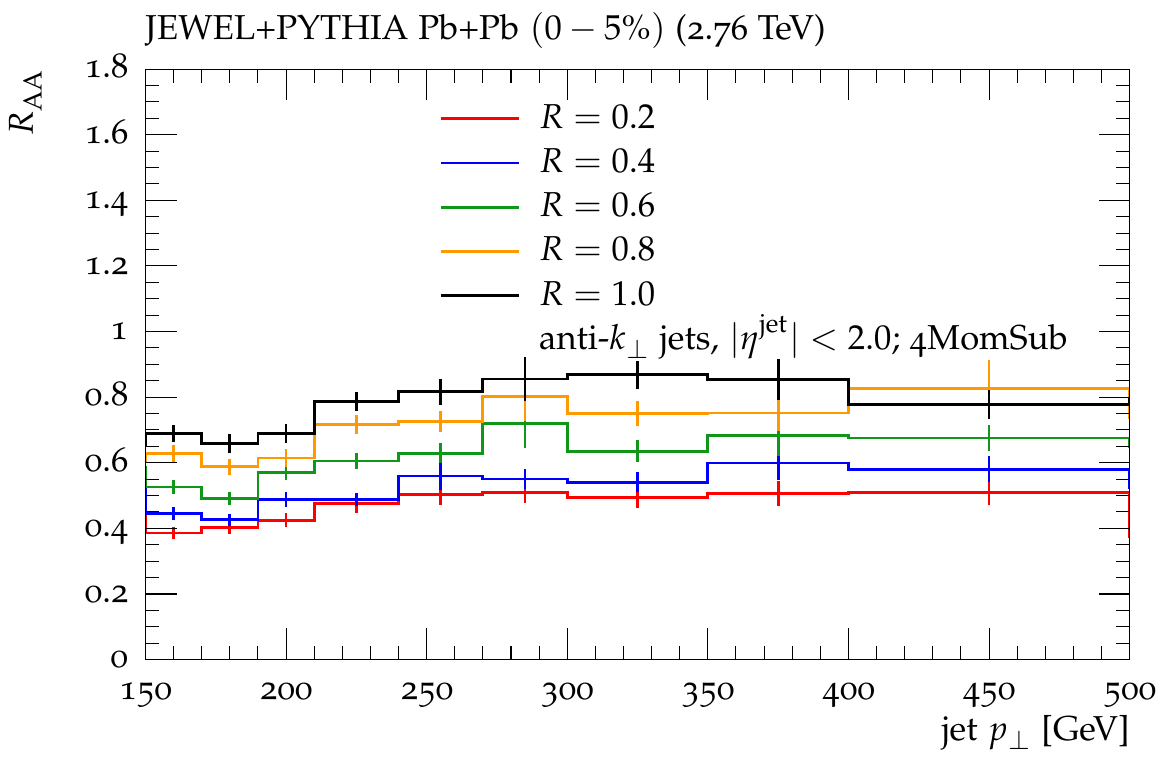} 
	\caption{Inclusive jet nuclear modification factors $R_\text{AA}$ in Pb+Pb central events in \textsc{Jewel+Pythia} for different jet radii $R$ and including medium response with 4MomSub.}
	\label{fig:RAAdifferentR}
\end{figure}		

The jet radius dependence of $R_\text{AA}$ is shown in Fig.~\ref{fig:RAAdifferentR} with medium response and 4MomSub. The expected increase of $R_\text{AA}$ with $R$, because with increasing jet radius more and more of the lost energy is recovered, is indeed observed\footnote{This is in contrast to the behaviour observed in~\cite{Casalderrey-Solana:2016jvj}, where $R_\text{AA}$ decreases with increasing jet radius because wider jets are more easily lost and medium response cannot compensate this loss.}.

\medskip

\begin{figure}[t]
	\centering
	\includegraphics[width=0.47\textwidth]{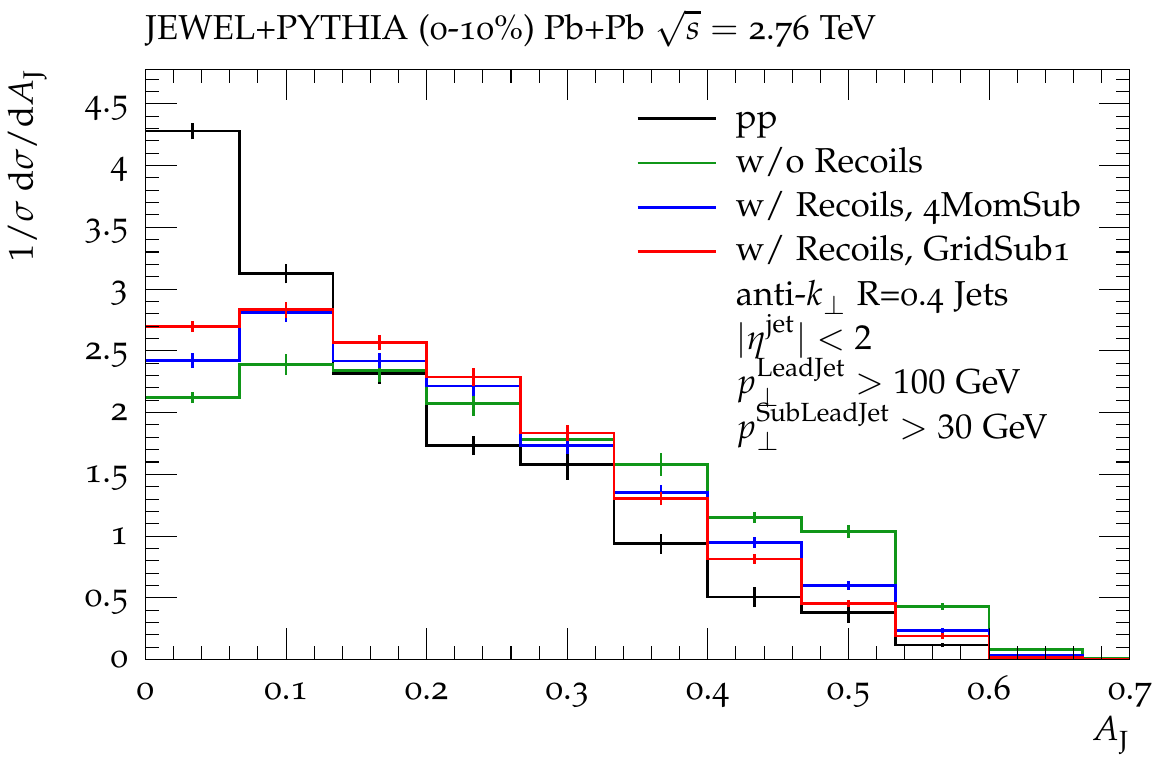} 
	\caption{Di-jet momentum asymmetry $A_J = (\pt^\text{LeadJet} - \pt^\text{SubLeadJet})/(\pt^\text{LeadJet} + \pt^\text{SubLeadJet})$ for central Pb+Pb central events in \textsc{Jewel+Pythia}. The green line represents \textsc{Jewel+Pythia} without medium response while the blue and red lines show the result including medium response with 4MomSub and GridSub1 respectively.}
	\label{fig:Aj}
\end{figure}		

\begin{figure}[t]
	\centering
	\includegraphics[width=0.47\textwidth]{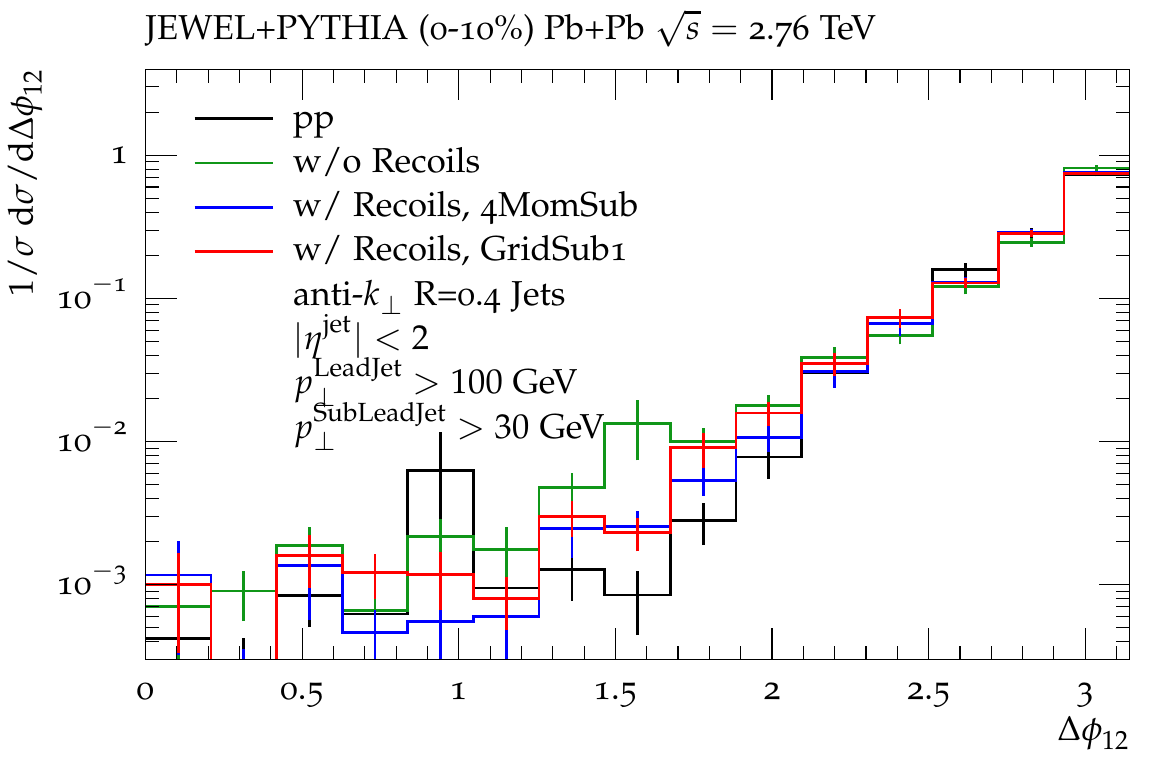} 
	\caption{Di-jet relative azimuthal angle $\Delta \phi_{12}$ for central Pb+Pb central events in \textsc{Jewel+Pythia}. The green line represents \textsc{Jewel+Pythia} without medium response while the blue and red lines show the result including medium response with 4MomSub and GridSub1 respectively.}
	\label{fig:Dphi}
\end{figure}		

Figs.~\ref{fig:Aj} and ~\ref{fig:Dphi} show the di-jet momentum asymmetry 
\begin{equation}
A_J = \frac{\pt^\text{LeadJet} - \pt^\text{SubLeadJet}}{\pt^\text{LeadJet} + \pt^\text{SubLeadJet}}
\end{equation}
and relative azimuthal angle $\Delta \phi_{12}$, respectively. Here, the leading jet is required to have $\pt^\text{LeadJet} > \unit[100]{GeV}$ and the cut on the sub-leading jet is $\pt^\text{SubLeadJet} > \unit[30]{GeV}$\footnote{Analysis cuts are always applied after subtraction.}. The momentum asymmetry $A_J$ is calculated without $\Delta \phi_{12}$ cut. The jet axis and thus $\Delta \phi_{12}$ are unaffected by medium response, while in the case of $A_J$ it leads to a mild reduction of the medium modification obtained without medium response. 

\section{Application to jet shape observables}
\label{sec::jetshapes}

In contrast to the observables discussed in the previous section, that aim at characterising global properties of jet events, jet shape observables are sensitive to the momentum distribution inside the jet. The latter are thus more affected by medium response. The energy in QCD jets is very much concentrated towards the jet axis, while medium response leads to a much broader distribution of relatively soft activity. Also the fluctuations of the two components are different. In this section we discuss a number of jet shape observables and how they are affected by medium response in \textsc{Jewel}.

\subsection{Jet mass}

\begin{figure}[h] 
	\centering
	\includegraphics[width=0.47\textwidth]{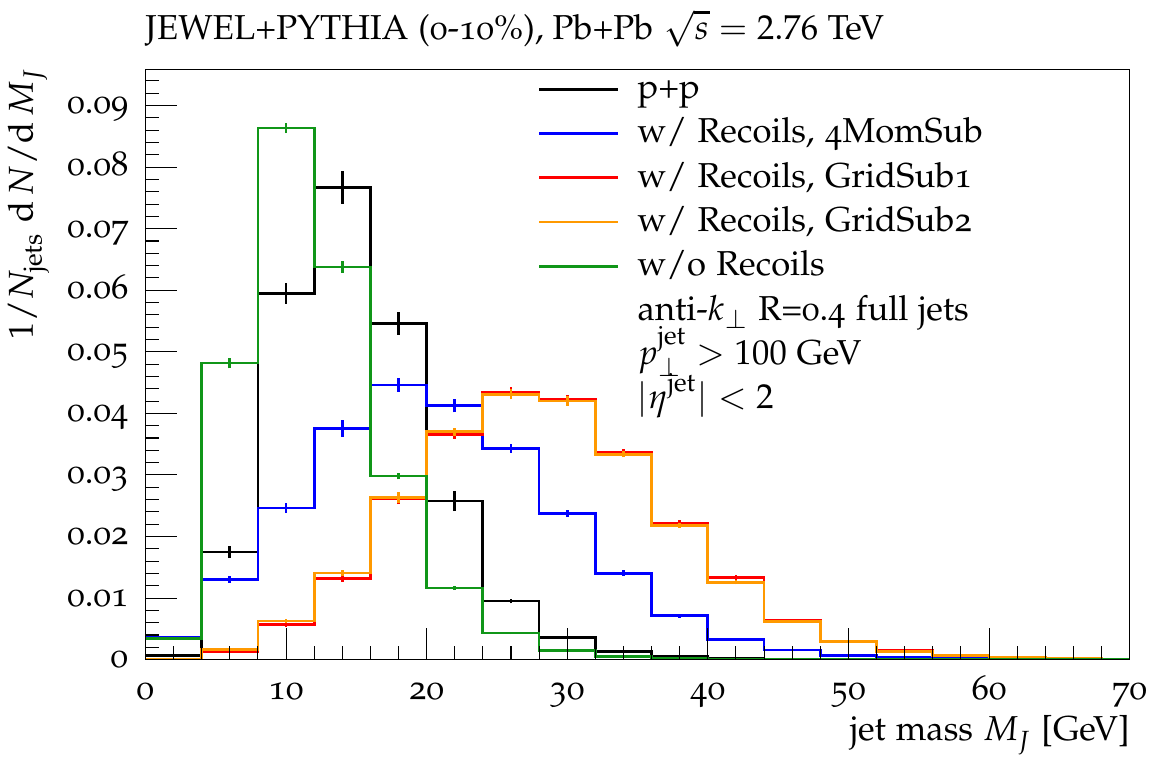} 
	\caption{Jet mass distributions in central Pb+Pb events for anti-$k_\perp$ full jets with radius parameter $R = 0.4$ and $\pt^\text{jet} > \unit[100]{GeV}$. The black line represents the mass in corresponding p+p collisions, while the green line is for \textsc{Jewel+Pythia} without medium response and the blue, red and orange lines correspond to \textsc{Jewel+Pythia} including medium response with 4MomSub, GridSub1 and GridSub2 subtraction, respectively.} 
	\label{fig:JetMass}
\end{figure}

\begin{figure}[h] 
	\centering
	\includegraphics[width=0.47\textwidth]{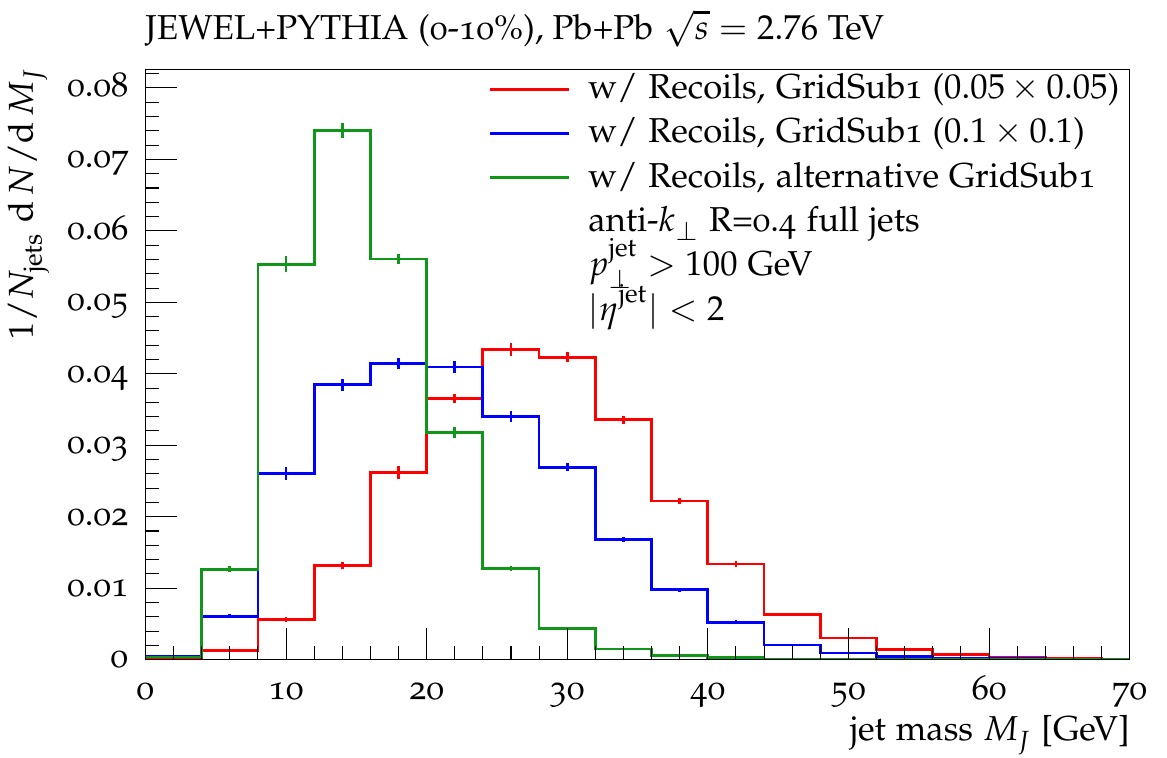} 
	\caption{Jet mass distributions in central Pb+Pb events for anti-$k_\perp$ full jets with radius parameter $R = 0.4$ and $\pt^\text{jet} > \unit[100]{GeV}$ with medium response and variations of the GridSub1 subtraction. The red histogram is the default (with cell size $0.05 \times 0.05$), in the blue the cell size is increased to $0.1 \times 0.1$, and the green is with default cell size but instead of four-momenta the energies of particles inside the cells are summed and the cell momentum is assumed to be massless.} 
	\label{fig:gridJetMass}
\end{figure}

\begin{figure*}[h] 
	\centering
	\includegraphics[width=0.47\textwidth]{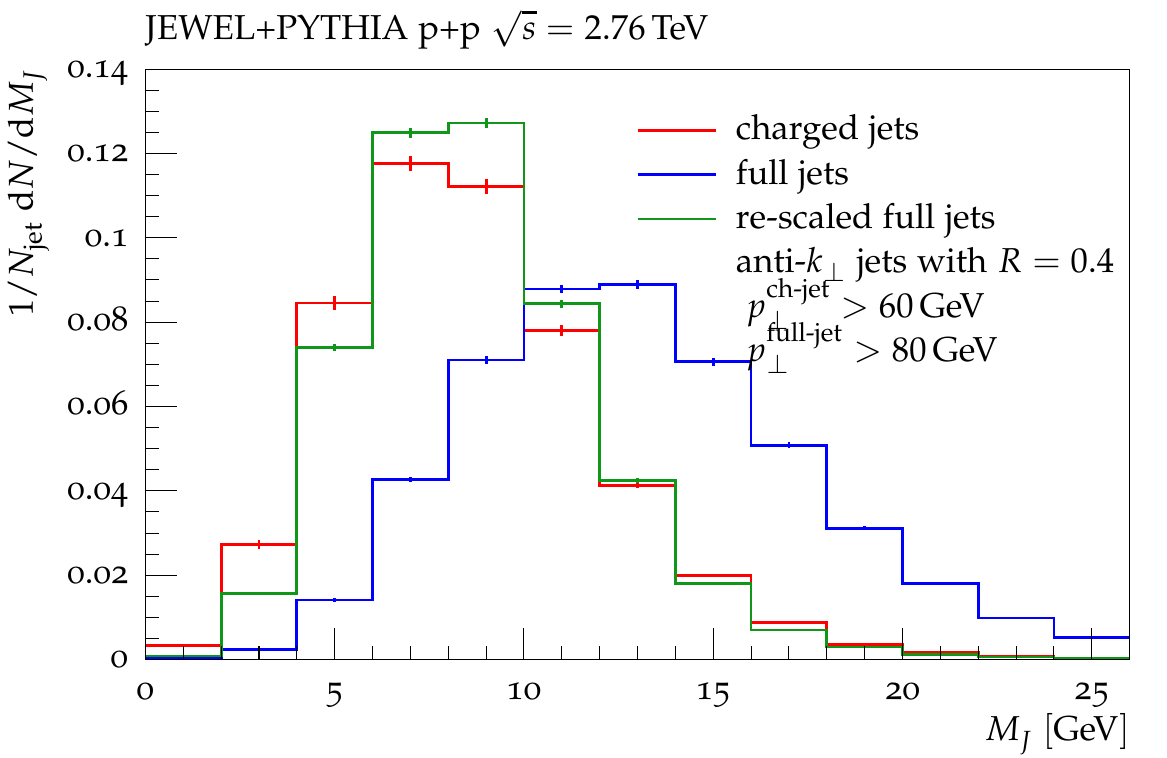} 
	\includegraphics[width=0.47\textwidth]{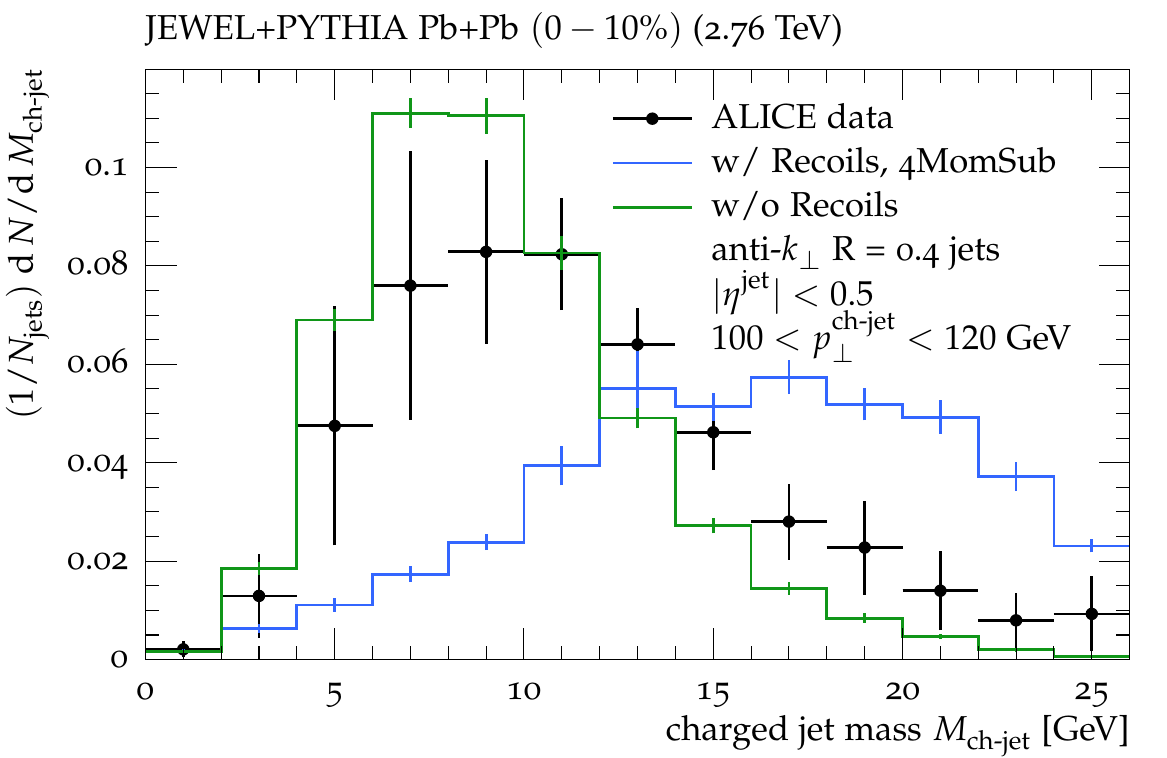} 
	\caption{Left: Jet mass distributions from \textsc{Jewel+Pythia} for p+p. The blue histogram shows the full jet distribution, the red the one for charged jets, and the green histogram is the re-scaled blue histogram. Right: Comparison of the re-scaled full jet mass distribution with recent ALICE data~\cite{Acharya:2017goa} for the charged jet mass.}
	\label{fig:aliceChJetMass}
\end{figure*}

The reconstructed jet mass is a good probe of medium induced jet modifications and medium response, since it is sensitive to the soft sector. Fig.~\ref{fig:JetMass} shows the \textsc{Jewel+Pythia} results for the jet mass distribution. The Monte Carlo shows a shift towards larger masses when medium response is included, whilst for events generated without recoils, a smaller jet mass is observed for jets belonging to the same kinematic range. The latter is due to the known narrowing of the hard jet core. The partial cancellation between two competing effects -- the narrowing due to energy loss and the broadening due to medium response -- is typical for this kind of observables and also seen in other jet shapes (e.g.\ the jet profile and girth). We observe a large difference between 4MomSub and GridSub subtraction in this observable, but good agreement between the two versions GridSub1 and GridSub2. In fact, the jet mass is very sensitive to the details of the grid subtraction. In Fig.~\ref{fig:gridJetMass} we compare two different cell sizes and two ways of computing the cell momentum. One is the default, which consists of summing the four-momenta of the particles in the cell (and subtracting the thermal momenta), and the other sums the particles' energies and assumes the cell four-momentum to be massless and to point in the direction defined by the cell centre. Both variations lead to large differences in the jet mass distribution (which is not observed in any other observable we studied). We therefore strongly discourage the use of grid subtraction for the jet mass and from here on show results only for 4MomSub subtraction.

As discussed in section~\ref{sec:limitations}, in order to be able to compare the \textsc{Jewel+Pythia} results to the ALICE data, the charged jet mass has to be estimated from the full jet mass. We do this by re-scaling the full jet mass with a constant factor 2/3 and the jet $\pt$ with a factor 3/4 (this is needed to match the $\pt$ cuts in the charged jet sample). The scaling factors are extracted from the \textsc{Jewel+Pythia} p+p sample. The left panel of Fig.~\ref{fig:aliceChJetMass} shows the charged jet, full jet and re-scaled full jet mass distributions in p+p and gives a lower bound on the related systematic uncertainties. We would like to stress once more that this is an ad hoc procedure and that there is no guarantee that it yields meaningful results. The right panel of Fig.~\ref{fig:aliceChJetMass} shows the comparison of the re-scaled full jet mass distribution from \textsc{Jewel+Pythia} to a recent ALICE measurement~\cite{Acharya:2017goa}. The Monte Carlo predicts significantly larger jet masses, but given the uncertainties involved in obtaining the charged jet distribution it is difficult to interpret this comparison with data.

\subsection{Fragmentation functions}

\begin{figure*}[h] 
	\centering
	\includegraphics[width=0.47\textwidth]{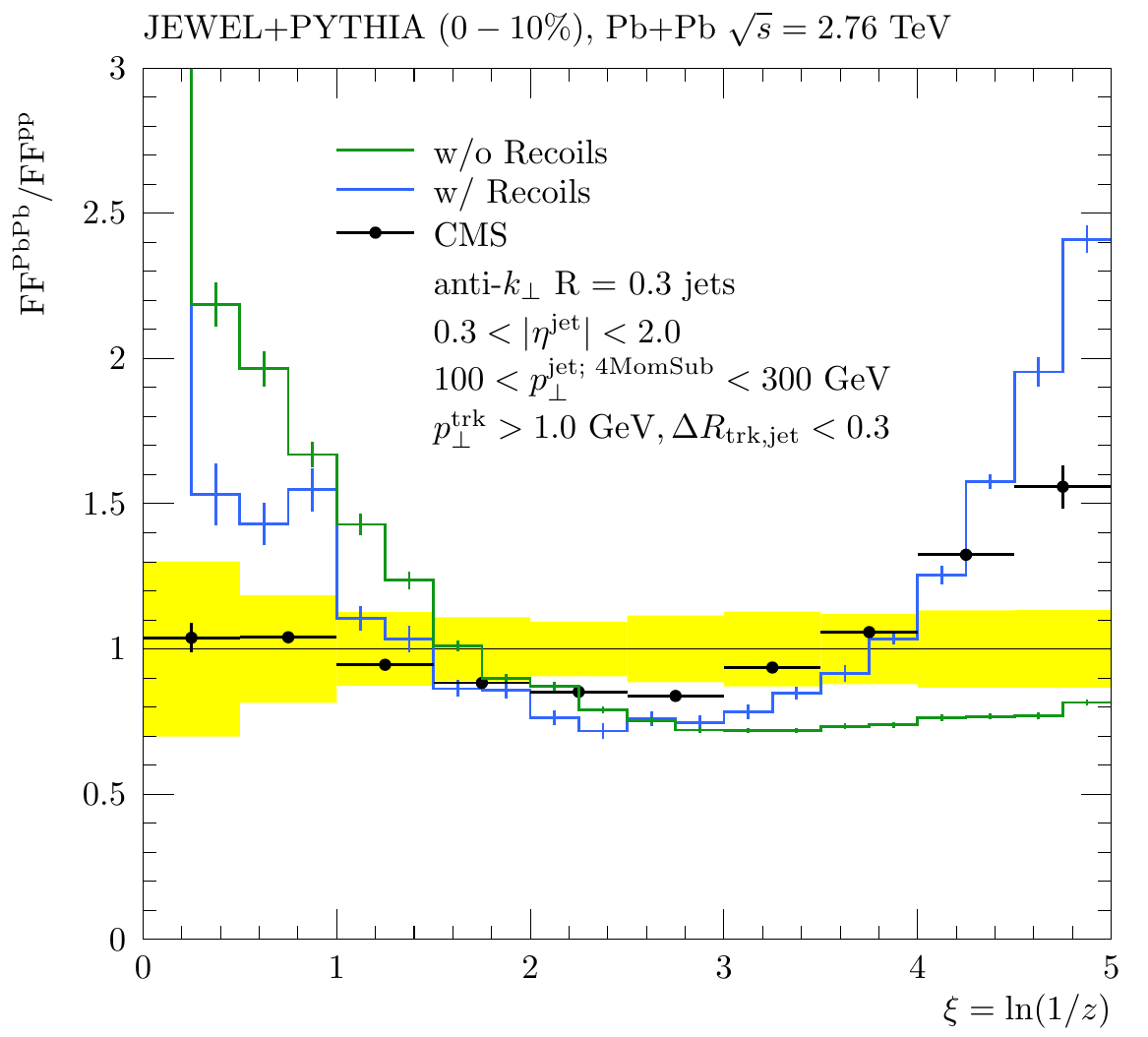} 
	\includegraphics[width=0.47\textwidth]{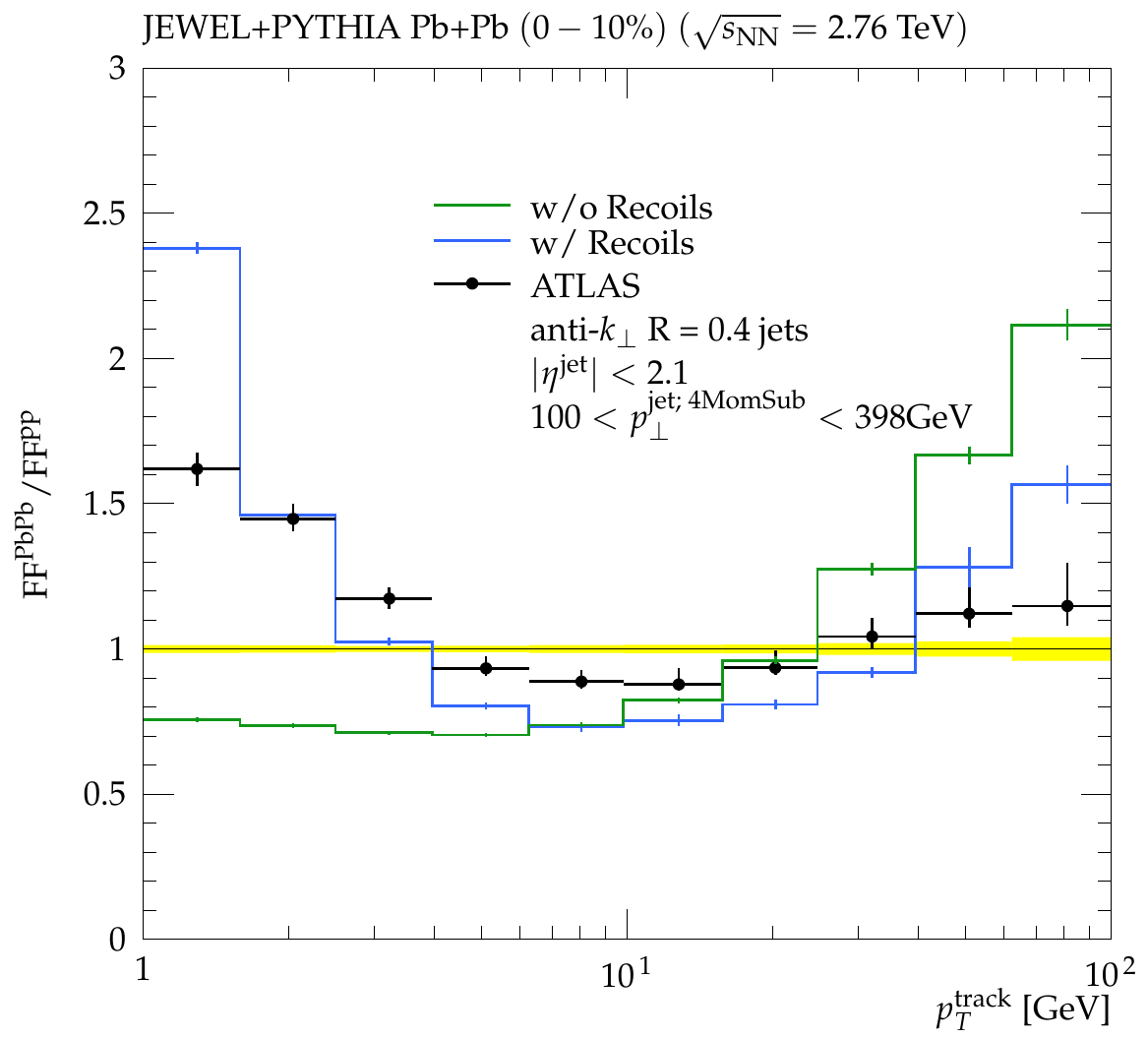} 
	\caption{The ratio of jet fragmentation functions (FF) Pb+Pb to p+p  compared with CMS~\cite{Chatrchyan:2014ava} (left) and ATLAS data~\cite{ATLAS-CONF-2015-055} (right). The data systematic uncertainties are shown in the yellow band around unity. Medium response is included in \textsc{Jewel+Pythia} results shown as blue histograms, but the subtraction (in this case 4MomSub) can only be applied to the jet $\pt$ and not to the tracks. The corresponding \textsc{Jewel+Pythia} results without medium response are shown as green histograms.}
	\label{fig:jewelFragFunction}
\end{figure*}

Intra-jet fragmentation function~\cite{Chatrchyan:2014ava,Aad:2014wha,ATLAS-CONF-2015-055} in p+p and Pb+Pb collisions are also an important jet sub-structure observable. However, in \textsc{Jewel} there is no way of doing the subtraction for individual hadrons or, as in this case, tracks. In Fig.~\ref{fig:jewelFragFunction}, which shows the modification of the fragmentation function in Pb+Pb collisions compared to p+p, we therefore in the sample with medium response correct the jet $\pt$, but all tracks enter the fragmentation function. It is thus expected that \textsc{Jewel+Pythia} overshoots the data in the low $z$ or $\pt$, corresponding to high $\xi$, region. The sample without medium response in this region shows a suppression as opposed to the enhancement seen in the data and the sample with recoiling partons, confirming the interpretation that the low $\pt$ (high $\xi$) enhancement seen in the data is due to medium response. The enhancement at high $\pt$ (low $\xi$) region is caused by the already mentioned narrowing and hardening of the hard jet core, and is more pronounced in \textsc{Jewel+Pythia} than in data. It is stronger without medium response, because the latter does not affect the hard fragments, but slightly increases the jet $\pt$.

\subsection{Jet profile}

\begin{figure}[h] 
	\centering
	\includegraphics[width=0.47\textwidth]{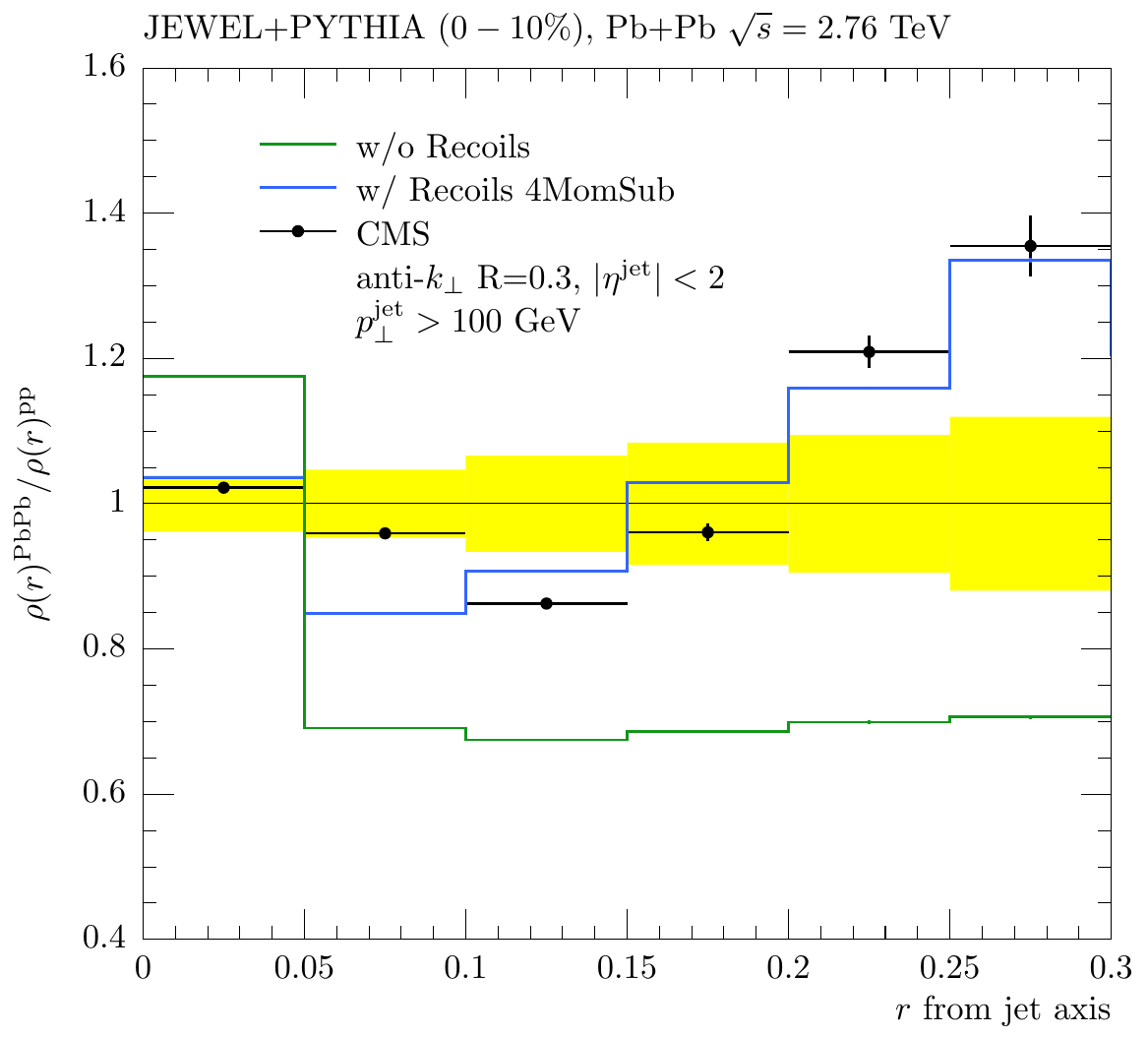} 
	\caption{Ration of the differential jet shape (or jet profile) in Pb+Pb and p+p measured by CMS~\cite{Chatrchyan:2013kwa} (black points) and compared with \textsc{Jewel+Pythia} results with (blue line) and without medium response (green line). The data systematic uncertainties are shown in the yellow band around unity. }
	\label{fig:cmsJetShape}
\end{figure}

The differential jet shape or jet profile $\rho(r)$ measures what fraction of the jet $\pt$ is found at what distance from the jet axis. It is defined as 
\begin{equation}
\rho(r) = \frac{1}{\pt^\text{jet}} \hspace{-2mm} \sum_{\substack{k \text{\ with\ }\\ \Delta R_{kJ} \in [r, r+\delta r]}} \hspace{-3mm} \pt^{(k)} \,,
\end{equation}
where the sum runs over all particles in the jet. The CMS measurement~\cite{Chatrchyan:2013kwa} was performed using the full jet $\pt$, but $\rho(r)$ was built only from tracks. Therefore, as is the case of the fragmentation function, we can do the subtraction for the jet $\pt$, but not for the charged particles. In this case, however, this is not a problem, since the jet profile built from tracks and the one built form all particles differ only by a constant factor. Assuming this factor to be the same in p+p and Pb+Pb, it will cancel exactly in the ratio of the jet profiles. We can therefore compare \textsc{Jewel+Pythia} results for full jets directly to the CMS data on the jet profile ratio. A more serious problem is that in experimental analysis only tracks with $\pt^\text{trk} > \unit[1]{GeV}$ are included. Since we can only subtract for the inclusive final state, this leads to a small mismatch, that becomes visible only at large $r$ and reaches up to \unit[10]{\%} in the highest $r$ bin.

Fig.~\ref{fig:cmsJetShape} shows the \textsc{Jewel+Pythia} result compared with CMS data~\cite{Chatrchyan:2013kwa} for the modification of the differential jet shape $\rho^{\text{PbPb}}/\rho^{\text{pp}}$ in Pb+Pb collisions compared to p+p. Including medium response and after performing the subtraction using the 4MomSub method, we are able to reproduce the general trend of the data. \textsc{Jewel+Pythia} with recoiling partons describes the enhancement of the jet shape at large radii mostly due to soft particles ($\pt < \unit[3]{GeV}$), while without medium response the enhancement is entirely absent. 

\subsection{Girth}

\begin{figure*}[h] 
	\centering
	\includegraphics[width=0.47\textwidth]{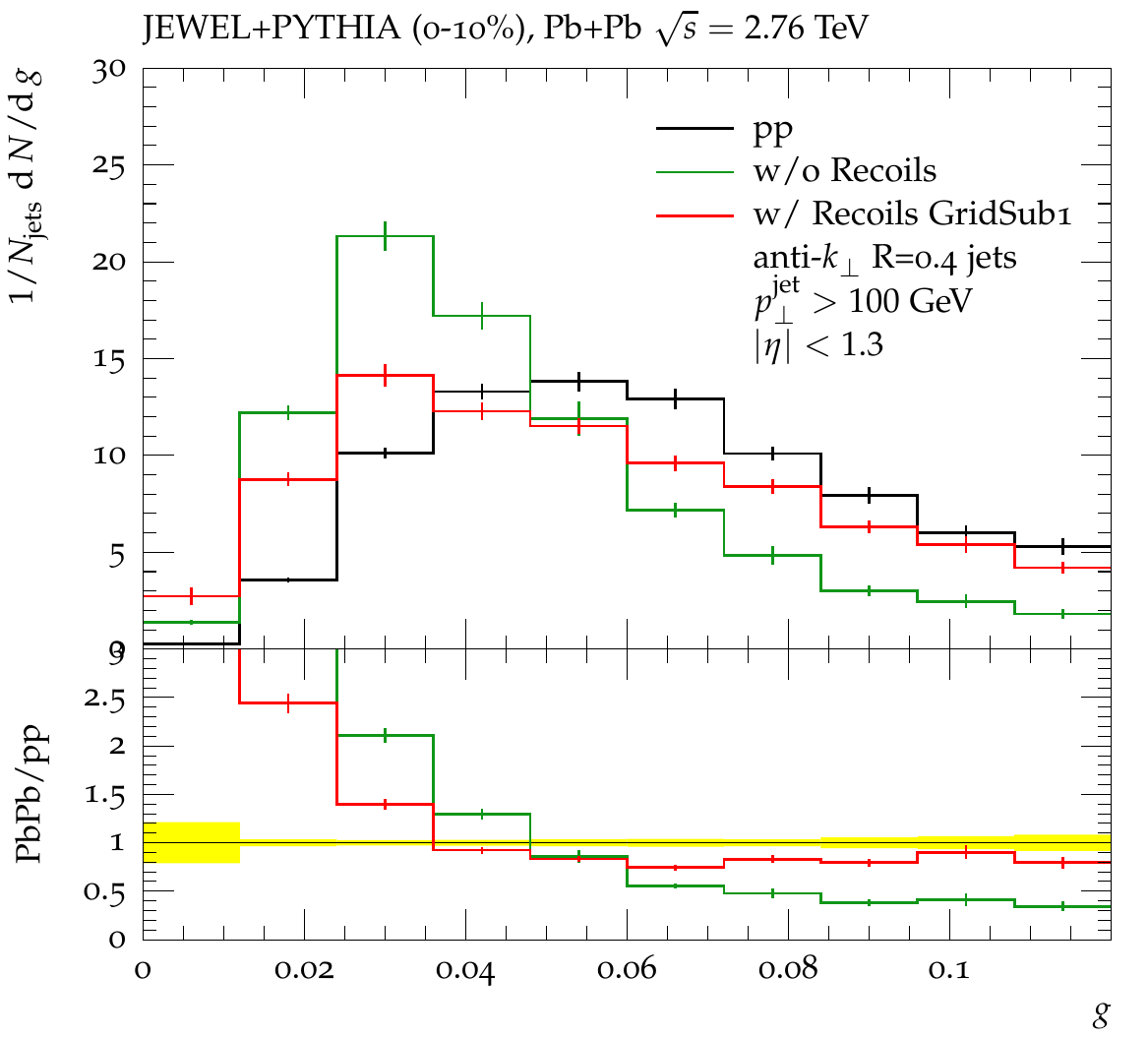} 
	\includegraphics[width=0.47\textwidth]{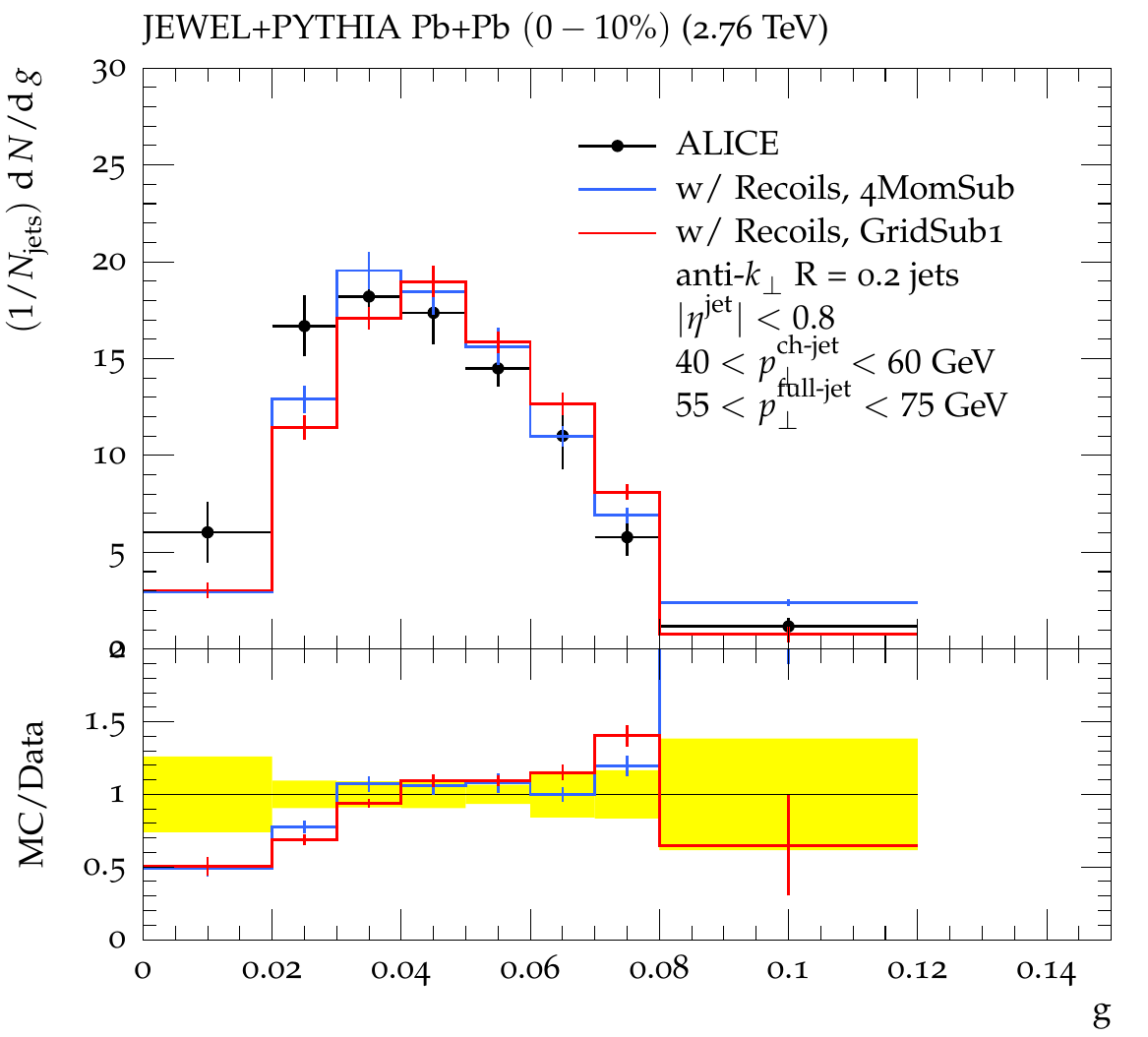} 
	\caption{Left: Distribution of the first radial moment (girth $g$) for $R=0.4$ fully reconstructed jets with $\pt^\text{jet} > \unit[100]{GeV}$ in central Pb+Pb collisions from \textsc{Jewel+Pythia}. The black histogram shows the corresponding p+p result, the green Pb+Pb without medium response and the red Pb+Pb including medium response with GridSub1 subtraction. The yellow shaded region around unity on the left panel highlights the statistical uncertainty in the p+p reference. Right: ALICE data~\cite{Cunqueiro:2015dmx} for charged jets ($R = 0.2$ and $\unit[40]{GeV} < \pt^\text{jet} < \unit[60]{GeV}$) compared with \textsc{Jewel+Pythia} for full jets (with adjusted $\pt$ range). The yellow shaded region around unity represents the data systematic uncertainties. }
	\label{fig:alicegirth}
\end{figure*}

The first radial moment of the jet profile is called girth~\cite{Giele:1997hd} and is defined as
\begin{equation}
g = \frac{1}{\pt^\text{jet}} \sum_{k\in J} \pt^{(k)} \Delta R_{kJ} \,,
\end{equation}
where the numerator sums the distance from the jet axis weighted with $\pt^{(k)}$ of each constituent $k$ of the jet. It characterises the width of the $\pt$ distribution inside the jet. 

\textsc{Jewel+Pythia} results for girth using GridSub1 subtraction for fully reconstructed jets in central Pb+Pb collisions are shown in the left panel of Fig.~\ref{fig:alicegirth}. We find a shift to smaller values of $g$ due to narrowing of the hard component, which is partly compensated by a broadening of the jet due to medium response. We also compare our results with preliminary ALICE data~\cite{Cunqueiro:2015dmx} for charged jets in the right panel of Fig.~\ref{fig:alicegirth}. Following the same argument as above for the jet profile, the girth of full and charged jets should be the same, provided the $\pt$ range is adjusted accordingly. We confirmed this in the Monte Carlo for p+p collisions. We therefore in Fig.~\ref{fig:alicegirth} compare \textsc{Jewel+Pythia} results for fully reconstructed jets at a correspondingly higher $\pt$ with the ALICE data. We find reasonable agreement, but the \textsc{Jewel+Pythia} distribution peaks at slightly higher values than the data.

\subsection{Groomed shared momentum fraction $z_g$}

\begin{figure*}[h] 
	\centering
	\includegraphics[width=0.47\textwidth]{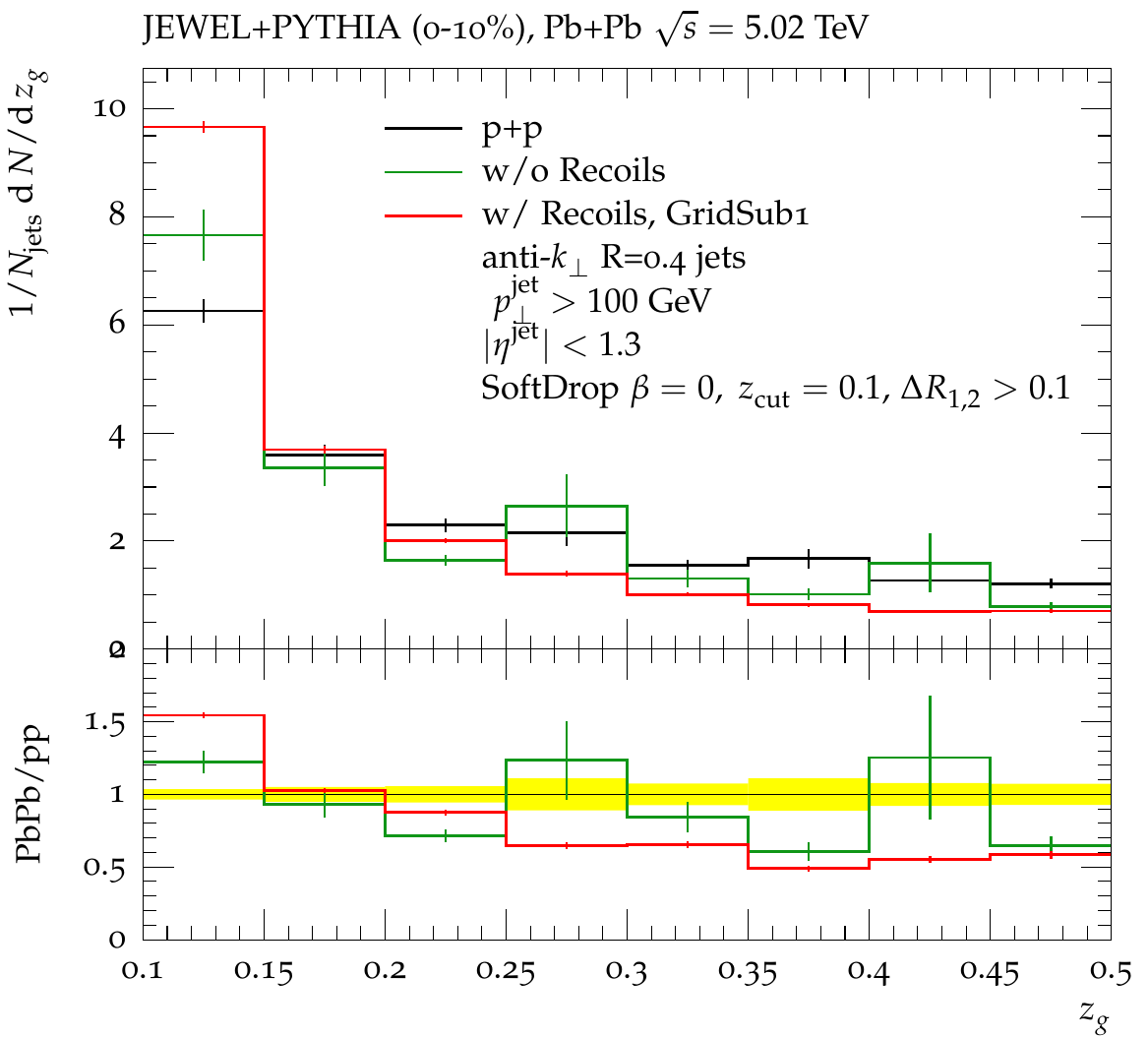} 
	\includegraphics[width=0.47\textwidth]{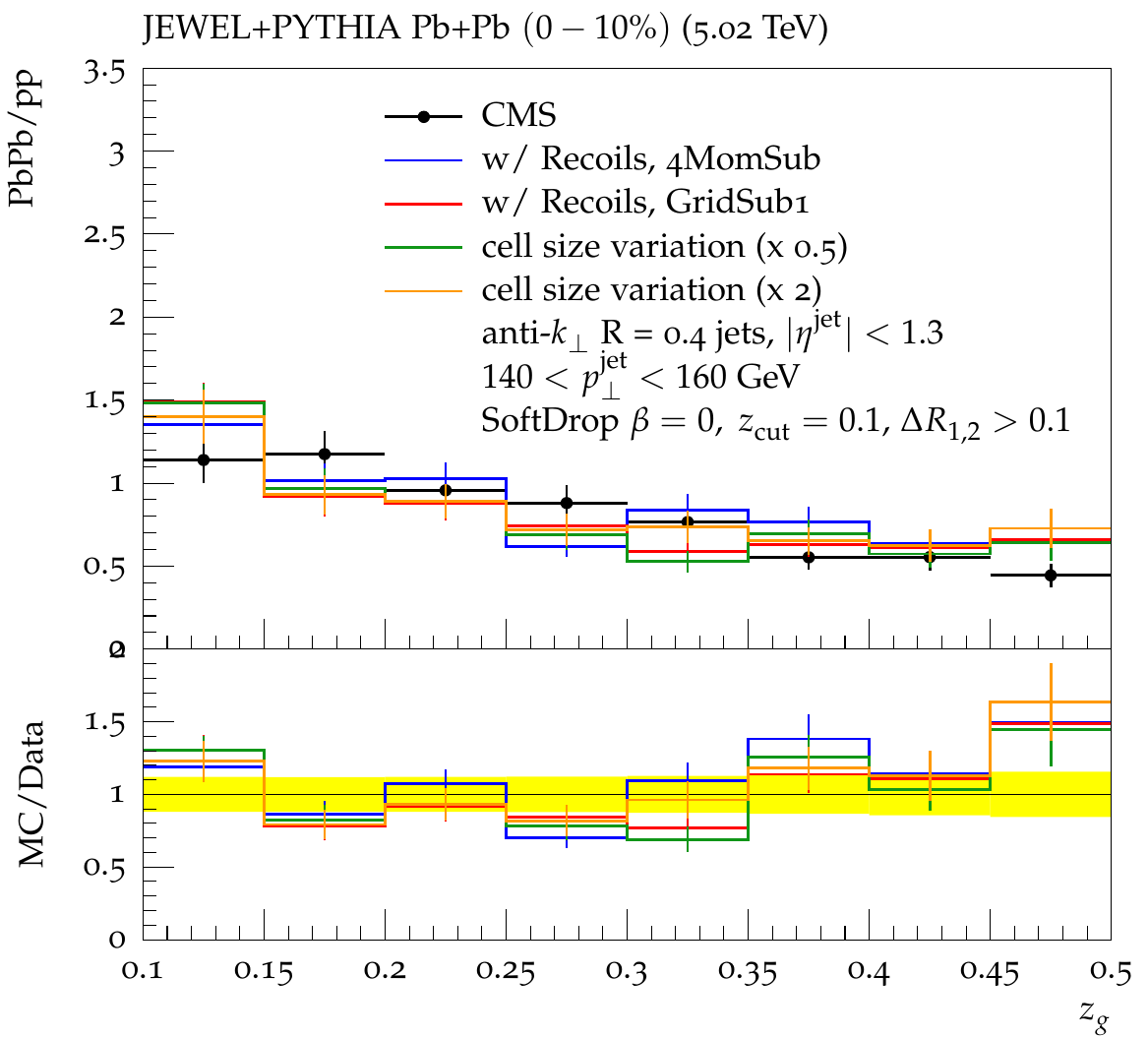} 
	\caption{\textsc{Jewel+Pythia} predictions for the groomed shared momentum fraction $z_g$ in central Pb+Pb events and p+p events. Left: $z_g$ distribution in p+p (black), central Pb+Pb collisions without recoiling partons (green) and with medium response and GridSub1 subtraction (red) for jets with $\pt^\text{jet} > \unit[100]{GeV}$ and Soft Drop parameters $z_\text{cut}=0.1$ and $\beta = 0$. Right: Comparison of \textsc{Jewel+Pythia} results with different grid sizes to CMS data~\cite{CMS:2016jys}. Note that the data is not unfolded, but the resolution is not published so no smearing is applied to the Monte Carlo events. A comparison to properly smeared \textsc{Jewel+Pythia} results can be found in~\cite{CMS:2016jys}. The yellow shaded region around unity in the left panel highlights the statistical uncertainty in the p+p reference and on the right represents the the data systematic uncertainties.}
	\label{fig:cmsSplitting}
\end{figure*}

The groomed shared momentum fraction $z_g$ is a measure for the momentum asymmetry in the hardest, i.e.\ largest angle, two-prong structure in the jet. In p+p collisions it is closely related to the Altarelli-Parisi splitting function~\cite{Larkoski:2015lea}. It is defined through the Soft Drop procedure~\cite{Dasgupta:2013ihk,Larkoski:2014wba} detailed below and implemented in \textsc{FastJet}~\cite{Cacciari:2011ma} contrib. First, jets are clustered with the anti-$k_\perp$ algorithm and re-clustered with Cambridge/Aachen. Then the last clustering step is undone, yielding the largest angle two-prong structure in the jet. If this configuration satisfies the Soft Drop condition
\begin{equation}
z_g = \frac{\text{min}(p_{\perp,1}, p_{\perp,2})}{p_{\perp,1}+ p_{\perp,2}} > z_{\rm{cut}} \left(\frac{\Delta R_{1,2}}{R_{J}}\right)^{\beta}
\end{equation}
where $z_{cut}$ and $\beta$ are parameters, it is kept. Otherwise, the softer of the two prongs is discarded and the procedure of un-doing the last clustering step is repeated for the harder prong. In this way soft contaminations are systematically removed from the jet and the hardest two-prong structure is identified. Soft Drop jet grooming thus takes an inclusive jet collection and turns it into a different collection of jets with two-prong structure of a minimum momentum symmetry provided by $z_{\rm{cut}}$.  Varying $z_{\rm{cut}}$ up or down varies the degree of asymmetrical splitting allowed in the  parton's fragmentation, while the $\beta$ controls how collinear the configuration has to be.

In p+p collisions, this method is has been studied in some detail~\cite{Larkoski:2014wba,Larkoski:2015lea}, but in heavy ion collisions the exact meaning of the grooming procedure is not obvious, due to the presence of the fluctuating underlying heavy ion event and the increased soft sector, that the procedure tries to remove. Recent analytical studies~\cite{Mehtar-Tani:2016aco} have shown that grooming increases the sensitivity to medium induced gluon bremsstrahlung thus experimentally opening up different avenues to directly probe the effect of the medium on a jet by jet basis. In \textsc{Jewel}, however, a different story unfolds. 

As shown in the right panel of Fig.~\ref{fig:cmsSplitting}, there is an increase in asymmetrical splittings in Pb+Pb jets as opposed to p+p jets, which is observed in recent preliminary CMS results and reproduced in \textsc{Jewel+Pythia}. The secondary feature observed in this measurement is an apparent reduction of the effect for higher $\pt$ jets. \textsc{Jewel} reproduces this behavior qualitatively as well, with very high $\pt$ jets showing very little difference in the momentum fraction of the first splitting~\cite{CMS:2016jys}. The left panel of Fig.~\ref{fig:cmsSplitting} shows that in the Monte Carlo the modification of the $z_g$ distribution in Pb+Pb collisions is partly due to the narrowing of the jet, as seen in the sample without medium response. The more important contribution, however, comes from adding the recoiling partons\footnote{For a detailed discussion of the origins of the effect in \textsc{Jewel} cf.~\cite{zgpaper}.}. In \textsc{Jewel+Pythia} we see no sign of medium induced bremsstrahlung contributing to the effect, as advertised in~\cite{Mehtar-Tani:2016aco}. 

\section{Discussion and conclusions}
\label{sec::conclusions}

Studies of jet sub-structure modifications in heavy ions probe the intricate interactions between the medium and jets. Due to their sensitivity to medium response, they offer the power to discriminate between several models and shed light on the underlying jet quenching mechanisms as well as the thermalisation of the deposited energy and momentum.

In \textsc{Jewel} it is possible to study medium response in detail by keeping the partons recoiling against interaction with the jet in the event. One has to keep in mind that this is only a limiting case, since these partons do not undergo further interactions in the medium. In order to be able to compare these results to experimental data, the thermal component of the recoiling partons' momenta has to be subtracted. In this paper we introduced two methods for doing this, a four-momentum and a grid based one. With these tools we can for the first time quantitatively study jet shape observables.

We find that -- at least in \textsc{Jewel+Pythia} -- a number of qualitative feature in the data can only be explained by medium response. These are
\begin{itemize}
	\item the increase at low $z$ of the ratio of intra-jet fragmentation functions in Pb+Pb compared to p+p,
	\item the increase of the jet profile at large distance from the jet axis in Pb+Pb compared to p+p,
	\item and the enhancement of asymmetric two-prong structures in Pb+Pb compared to p+p as seen in the $z_g$ distribution.
\end{itemize} 
This is in line with observations by other authors~\cite{Casalderrey-Solana:2016jvj,Tachibana:2017syd}. In other observables, in particular the jet mass and girth, a non-trivial cancellation between a narrowing of the jet core due to energy loss~\cite{Milhano:2015mng,Rajagopal:2016uip,Casalderrey-Solana:2016jvj} and a broadening due to medium response takes place. Also in the case of girth, including medium response leads to an improvement of the agreement between \textsc{Jewel+Pythia} and ALICE data. 

For the jet mass we find that the grid based subtraction does not yield reliable results. The 4MomSub subtration should be more robust, but without grid subtraction we do not have an independent way of cross-checking the results. We therefore recommend not to use GridSub for the jet mass and to take the comparison of \textsc{Jewel+Pythia} results to the ALICE data with a grain of salt.

Jet shape observables open a new perspective on jet quenching and may also help to address the question of thermalisation, and it is important to develop tools capable of quantitatively describing medium response. The present study with \textsc{Jewel} can only be a first step in this direction. As emphasised above, the treatment of recoiling partons is still schematic. The subtraction methods introduced in this paper are solid, but have their limitations, in particular when it comes to the description of charged jets. It is currently also impossible to perform the subtraction for particles (for instance in the fragmentation functions), due to the mix of parton and hadron level in the subtraction. The grid method also introduces systematic uncertainties related to the discretisation, that can, however, be quantified (cf. section~\ref{sec::systematics}). Nevertheless, the results for jet shapes obtained with \textsc{Jewel+Pythia} are very promising. In some cases this is the first time that they can be studied quantitatively in a consistent jet quenching model including medium response.

Upcoming measurements at the LHC will further advance the understanding of jet shapes by utilizing the jet grooming tools, amongst others. This ushers in a new era of sub-structure studies in heavy ion collisions, where correlation between different observables could point the way to the future in decoupling several of the physics features hidden in individual observables.

\acknowledgments

We are grateful to Guilherme Milhano for helpful discussions and comments on the manuscript. 
This work was supported by Funda\c{c}\~{a}o para a Ci\^{e}ncia e a Tecnologia (Portugal) under project CERN/FIS-NUC/0049/2015 and contract `Investigador FCT - Starting Grant' IF/00634/2015 (KCZ) and by the European Union as part of the FP7 Marie Curie Initial Training Network MCnetITN (PITN-GA-2012-315877) (RKE). RKE also acknowledges support from the National Science Foundation under Grant No.1067907 \& 1352081.


\begin{thebibliography}{99}

\bibitem{Aad:2014bxa} G.~Aad {\it et al.} [ATLAS Collaboration],
Phys.\ Rev.\ Lett.\  {\bf 114} (2015) no.7,  072302
doi:10.1103/PhysRevLett.114.072302
[arXiv:1411.2357 [hep-ex]].

\bibitem{Adam:2015ewa} J.~Adam {\it et al.} [ALICE Collaboration],
Phys.\ Lett.\ B {\bf 746} (2015) 1
doi:10.1016/j.physletb.2015.04.039
[arXiv:1502.01689 [nucl-ex]].

\bibitem{Khachatryan:2016jfl} V.~Khachatryan {\it et al.} [CMS Collaboration],
arXiv:1609.05383 [nucl-ex].

\bibitem{Khachatryan:2015lha} V.~Khachatryan {\it et al.} [CMS Collaboration],
JHEP {\bf 1601} (2016) 006
doi:10.1007/JHEP01(2016)006
[arXiv:1509.09029 [nucl-ex]].

\bibitem{Chatrchyan:2014ava} S.~Chatrchyan {\it et al.} [CMS Collaboration],
Phys.\ Rev.\ C {\bf 90}, no. 2, 024908 (2014)
doi:10.1103/PhysRevC.90.024908
[arXiv:1406.0932 [nucl-ex]].

\bibitem{Aad:2014wha} G.~Aad {\it et al.} [ATLAS Collaboration],
Phys.\ Lett.\ B {\bf 739} (2014) 320
doi:10.1016/j.physletb.2014.10.065
[arXiv:1406.2979 [hep-ex]].

\bibitem{ATLAS-CONF-2015-055} The ATLAS collaboration,
ATLAS-CONF-2015-055.

\bibitem{Chatrchyan:2013kwa} S.~Chatrchyan {\it et al.} [CMS Collaboration],
Phys.\ Lett.\ B {\bf 730}, 243 (2014)
doi:10.1016/j.physletb.2014.01.042
[arXiv:1310.0878 [nucl-ex]].

\bibitem{Khachatryan:2016tfj} V.~Khachatryan {\it et al.} [CMS Collaboration],
JHEP {\bf 1611} (2016) 055
doi:10.1007/JHEP11(2016)055
[arXiv:1609.02466 [nucl-ex]].

\bibitem{Cunqueiro:2015dmx} L.~Cunqueiro [ALICE Collaboration],
arXiv:1512.07882 [nucl-ex].

\bibitem{CMS:2016jys} CMS Collaboration [CMS Collaboration],
CMS-PAS-HIN-16-006.

\bibitem{Acharya:2017goa} S.~Acharya {\it et al.} [ALICE Collaboration],
arXiv:1702.00804 [nucl-ex].

\bibitem{Milhano:2015mng} J.~G.~Milhano and K.~C.~Zapp,
Eur.\ Phys.\ J.\ C {\bf 76} (2016) no.5,  288
doi:10.1140/epjc/s10052-016-4130-9
[arXiv:1512.08107 [hep-ph]].

\bibitem{Rajagopal:2016uip} K.~Rajagopal, A.~V.~Sadofyev and W.~van der Schee,
Phys.\ Rev.\ Lett.\  {\bf 116} (2016) no.21,  211603
doi:10.1103/PhysRevLett.116.211603
[arXiv:1602.04187 [nucl-th]].

\bibitem{Casalderrey-Solana:2016jvj} J.~Casalderrey-Solana, D.~Gulhan, G.~Milhano, D.~Pablos and K.~Rajagopal,
JHEP {\bf 1703} (2017) 135
doi:10.1007/JHEP03(2017)135
[arXiv:1609.05842 [hep-ph]].

\bibitem{KunnawalkamElayavalli:2016dda} R.~Kunnawalkam Elayavalli,
J.\ Phys.\ Conf.\ Ser.\  {\bf 832} (2017) no.1,  012004
doi:10.1088/1742-6596/832/1/012004
[arXiv:1610.09364 [nucl-th]].

\bibitem{Tachibana:2017syd} Y.~Tachibana, N.~B.~Chang and G.~Y.~Qin,
Phys.\ Rev.\ C {\bf 95} (2017) no.4,  044909
doi:10.1103/PhysRevC.95.044909
[arXiv:1701.07951 [nucl-th]].

\bibitem{Chien:2015hda} Y.~T.~Chien and I.~Vitev,
JHEP {\bf 1605} (2016) 023
doi:10.1007/JHEP05(2016)023
[arXiv:1509.07257 [hep-ph]].

\bibitem{Gubser:2007ga} S.~S.~Gubser, S.~S.~Pufu and A.~Yarom,
Phys.\ Rev.\ Lett.\  {\bf 100} (2008) 012301
doi:10.1103/PhysRevLett.100.012301
[arXiv:0706.4307 [hep-th]].

\bibitem{Chesler:2007sv} P.~M.~Chesler and L.~G.~Yaffe,
Phys.\ Rev.\ D {\bf 78} (2008) 045013
doi:10.1103/PhysRevD.78.045013
[arXiv:0712.0050 [hep-th]].

\bibitem{Chesler:2015lsa} P.~M.~Chesler and W.~van der Schee,
Int.\ J.\ Mod.\ Phys.\ E {\bf 24} (2015) no.10,  1530011
doi:10.1142/S0218301315300118
[arXiv:1501.04952 [nucl-th]].

\bibitem{CasalderreySolana:2011us} J.~Casalderrey-Solana, H.~Liu, D.~Mateos, K.~Rajagopal and U.~A.~Wiedemann,
book:Gauge/String Duality, Hot QCD and Heavy Ion Collisions. Cambridge, UK: Cambridge University Press, 2014
doi:10.1017/CBO9781139136747
[arXiv:1101.0618 [hep-th]].

\bibitem{Qin:2009uh} G.-Y.~Qin, A.~Majumder, H.~Song and U.~Heinz,
Phys.\ Rev.\ Lett.\  {\bf 103} (2009) 152303
doi:10.1103/PhysRevLett.103.152303
[arXiv:0903.2255 [nucl-th]].

\bibitem{Neufeld:2009ep} R.~B.~Neufeld and B.~Muller,
Phys.\ Rev.\ Lett.\  {\bf 103} (2009) 042301
doi:10.1103/PhysRevLett.103.042301
[arXiv:0902.2950 [nucl-th]].

\bibitem{Iancu:2015uja} E.~Iancu and B.~Wu,
JHEP {\bf 1510} (2015) 155
doi:10.1007/JHEP10(2015)155
[arXiv:1506.07871 [hep-ph]].

\bibitem{Bouras:2014rea} I.~Bouras, B.~Betz, Z.~Xu and C.~Greiner,
Phys.\ Rev.\ C {\bf 90} (2014) no.2,  024904
doi:10.1103/PhysRevC.90.024904
[arXiv:1401.3019 [hep-ph]].

\bibitem{He:2015pra} Y.~He, T.~Luo, X.~N.~Wang and Y.~Zhu,
Phys.\ Rev.\ C {\bf 91}, 054908 (2015)
doi:10.1103/PhysRevC.91.054908
[arXiv:1503.03313 [nucl-th]].

\bibitem{Gao:2016ldo} Z.~Gao, G.~L.~Ma, G.~Y.~Qin and H.~Z.~Zhang,
arXiv:1612.02548 [hep-ph].

\bibitem{Chen:2017zte} W.~Chen, S.~Cao, T.~Luo, L.~G.~Pang and X.~N.~Wang,
arXiv:1704.03648 [nucl-th].


\bibitem{Zapp:2013vla} K.~C.~Zapp,
Eur.\ Phys.\ J.\ C {\bf 74}, no. 2, 2762 (2014)
doi:10.1140/epjc/s10052-014-2762-1
[arXiv:1311.0048 [hep-ph]].


\bibitem{Floerchinger:2014yqa} S.~Floerchinger and K.~C.~Zapp,
Eur.\ Phys.\ J.\ C {\bf 74} (2014) no.12,  3189
doi:10.1140/epjc/s10052-014-3189-4
[arXiv:1407.1782 [hep-ph]].

\bibitem{Zapp:2012ak} K.~C.~Zapp, F.~Krauss and U.~A.~Wiedemann,
JHEP {\bf 1303} (2013) 080
[arXiv:1212.1599 [hep-ph]].

\bibitem{Zapp:2013zya} K.~C.~Zapp,
Phys.\ Lett.\ B {\bf 735} (2014) 157
[arXiv:1312.5536 [hep-ph]].

\bibitem{Zapp:2011ya} K.~C.~Zapp, J.~Stachel and U.~A.~Wiedemann,
JHEP {\bf 1107} (2011) 118
[arXiv:1103.6252 [hep-ph]].

\bibitem{Buckley:2010ar} A.~Buckley, J.~Butterworth, L.~Lonnblad, D.~Grellscheid, H.~Hoeth, J.~Monk, H.~Schulz and F.~Siegert,
Comput.\ Phys.\ Commun.\  {\bf 184} (2013) 2803
[arXiv:1003.0694 [hep-ph]].

\bibitem{Cacciari:2011ma} M.~Cacciari, G.~P.~Salam and G.~Soyez,
Eur.\ Phys.\ J.\ C {\bf 72}, 1896 (2012)
doi:10.1140/epjc/s10052-012-1896-2
[arXiv:1111.6097 [hep-ph]].

\bibitem{Abelev:2013kqa} B.~Abelev {\it et al.} [ALICE Collaboration],
JHEP {\bf 1403} (2014) 013
doi:10.1007/JHEP03(2014)013
[arXiv:1311.0633 [nucl-ex]].

\bibitem{Shen:2012vn} C.~Shen and U.~Heinz,
Phys.\ Rev.\ C {\bf 85} (2012) 054902
[Phys.\ Rev.\ C {\bf 86} (2012) 049903]
[arXiv:1202.6620 [nucl-th]].

\bibitem{Shen:2014vra} C.~Shen, Z.~Qiu, H.~Song, J.~Bernhard, S.~Bass and U.~Heinz,
Comput.\ Phys.\ Commun.\  {\bf 199}, 61 (2016)
doi:10.1016/j.cpc.2015.08.039
[arXiv:1409.8164 [nucl-th]].

\bibitem{Pumplin:2002vw} J.~Pumplin, D.~R.~Stump, J.~Huston, H.~L.~Lai, P.~M.~Nadolsky and W.~K.~Tung,
JHEP {\bf 0207} (2002) 012
doi:10.1088/1126-6708/2002/07/012
[hep-ph/0201195].

\bibitem{Eskola:2009uj} K.~J.~Eskola, H.~Paukkunen and C.~A.~Salgado,
JHEP {\bf 0904} (2009) 065
doi:10.1088/1126-6708/2009/04/065
[arXiv:0902.4154 [hep-ph]].

\bibitem{Whalley:2005nh} M.~R.~Whalley, D.~Bourilkov and R.~C.~Group,
hep-ph/0508110.

\bibitem{Cacciari:2008gp} M.~Cacciari, G.~P.~Salam and G.~Soyez,
JHEP {\bf 0804} (2008) 063
[arXiv:0802.1189 [hep-ph]].

\bibitem{Chatrchyan:2011ds} S.~Chatrchyan {\it et al.} [CMS Collaboration],
JINST {\bf 6}, P11002 (2011)
doi:10.1088/1748-0221/6/11/P11002
[arXiv:1107.4277 [physics.ins-det]].

\bibitem{Berta:2016ukt} P.~Berta [ATLAS Collaboration],
Nucl.\ Part.\ Phys.\ Proc.\  {\bf 273-275}, 1121 (2016).
doi:10.1016/j.nuclphysbps.2015.09.176

\bibitem{Aaboud:2017bzv} M.~Aaboud {\it et al.} [ATLAS Collaboration],
Eur.\ Phys.\ J.\ C {\bf 77} (2017) no.6,  379
doi:10.1140/epjc/s10052-017-4915-5
[arXiv:1702.00674 [hep-ex]].

\bibitem{Kodolova:2007hd} O.~Kodolova, I.~Vardanian, A.~Nikitenko and A.~Oulianov,
Eur.\ Phys.\ J.\ C {\bf 50} (2007) 117.
doi:10.1140/epjc/s10052-007-0223-9

\bibitem{Giele:1997hd} W.~T.~Giele, E.~W.~N.~Glover and D.~A.~Kosower,
Phys.\ Rev.\ D {\bf 57}, 1878 (1998)
doi:10.1103/PhysRevD.57.1878
[hep-ph/9706210].

\bibitem{Larkoski:2015lea} A.~J.~Larkoski, S.~Marzani and J.~Thaler,
Phys.\ Rev.\ D {\bf 91}, no. 11, 111501 (2015)
doi:10.1103/PhysRevD.91.111501
[arXiv:1502.01719 [hep-ph]].

\bibitem{Dasgupta:2013ihk} M.~Dasgupta, A.~Fregoso, S.~Marzani and G.~P.~Salam,
JHEP {\bf 1309} (2013) 029
doi:10.1007/JHEP09(2013)029
[arXiv:1307.0007 [hep-ph]].

\bibitem{Larkoski:2014wba} A.~J.~Larkoski, S.~Marzani, G.~Soyez and J.~Thaler,
JHEP {\bf 1405} (2014) 146
doi:10.1007/JHEP05(2014)146
[arXiv:1402.2657 [hep-ph]].

\bibitem{Mehtar-Tani:2016aco} Y.~Mehtar-Tani and K.~Tywoniuk,
arXiv:1610.08930 [hep-ph].

\bibitem{zgpaper} J.~G.~Milhano, U.~A.~Wiedemann and K.~C.~Zapp,
in preparation

\bibitem{Bjorken:1982tu} J.~D.~Bjorken,
FERMILAB-PUB-82-059-THY, FERMILAB-PUB-82-059-T.

\bibitem{Sjostrand:2006za} T.~Sjostrand, S.~Mrenna and P.~Z.~Skands,
JHEP {\bf 0605}, 026 (2006)
doi:10.1088/1126-6708/2006/05/026
[hep-ph/0603175].

\bibitem{Casalderrey-Solana:2015vaa} J.~Casalderrey-Solana, D.~C.~Gulhan, J.~G.~Milhano, D.~Pablos and K.~Rajagopal,
JHEP {\bf 1603} (2016) 053
doi:10.1007/JHEP03(2016)053
[arXiv:1508.00815 [hep-ph]].

\bibitem{Mehtar-Tani:2016xwr} Y.~Mehtar-Tani,
arXiv:1602.01047 [hep-ph].

\bibitem{KunnawalkamElayavalli:2016ttl} R.~Kunnawalkam Elayavalli and K.~C.~Zapp,
Eur.\ Phys.\ J.\ C {\bf 76} (2016) no.12,  695
doi:10.1140/epjc/s10052-016-4534-6
[arXiv:1608.03099 [hep-ph]].


\end{thebibliography}
\end{document}